\documentclass{IEEEtran}
\usepackage{amsmath,amssymb,amsfonts}
\usepackage{algorithmic}
\usepackage{graphicx,color}
\usepackage{textcomp}
\usepackage[caption=false]{subfig}
\usepackage{lipsum}
\usepackage{multirow}
\def\BibTeX{{\rm B\kern-.05em{\sc i\kern-.025em b}\kern-.08em
    T\kern-.1667em\lower.7ex\hbox{E}\kern-.125emX}}
\AtBeginDocument{\definecolor{ojcolor}{cmyk}{0.93,0.59,0.15,0.02}}

\def\b0{{\pmb{0}}} 

\def\ba{{\mathbf{a}}} \def\bb{{\mathbf{b}}}  
 \def\bff{{\mathbf{f}}} \def\bg{{\mathbf{g}}} \def\bh{{\mathbf{h}}}

  \def\bw{{\mathbf{w}}} \def\bx{{\mathbf{x}}}
\def\by{{\mathbf{y}}}   

\def\bA{{\mathbf{A}}}   
 \def\bF{{\mathbf{F}}} \def\bG{{\mathbf{G}}} \def\bH{{\mathbf{H}}}
\def\bI{{\mathbf{I}}}

  \def\bW{{\mathbf{W}}} 
\def\bY{{\mathbf{Y}}} 

\begin{document}

\title{Machine Learning-based Channel Prediction in Wideband Massive MIMO Systems with Small Overhead for Online Training}

\author{}
 \author{Beomsoo Ko, Hwanjin Kim, Minje Kim, and Junil Choi

\thanks{
         Beomsoo Ko, Minje Kim, and Junil Choi are with the School of Electrical Engineering, Korea Advanced Institute of Science and Technology, Daejeon 34141, South Korea
         (e-mail: kobs0318@kaist.ac.kr; mjkim97@kaist.ac.kr; junil@kaist.ac.kr).
         
         Hwanjin Kim is with Elmore Family School of Electrical and Computer Engineering, Purdue University, West Lafayette, IN 47907 USA (e-mail: kim4466@purdue.edu).   
      }
      
   }
   \maketitle

\begin{abstract}
Channel prediction compensates for outdated channel state information in multiple-input multiple-output (MIMO) systems. Machine learning (ML) techniques have recently been implemented to design channel predictors by leveraging the temporal correlation of wireless channels. However, most ML-based channel prediction techniques have only considered offline training when generating channel predictors, which can result in poor performance when encountering channel environments different from the ones they were trained on. To ensure prediction performance in varying channel conditions, we propose an online re-training framework that trains the channel predictor from scratch to effectively capture and respond to changes in the wireless environment. The training time includes data collection time and neural network training time, and should be minimized for practical channel predictors. To reduce the training time, especially data collection time, we propose a novel ML-based channel prediction technique called aggregated learning (AL) approach for wideband massive MIMO systems. In the proposed AL approach, the training data can be split and aggregated either in an array domain or frequency domain, which are the channel domains of MIMO-OFDM systems. This processing can significantly reduce the time for data collection. Our numerical results show that the AL approach even improves channel prediction performance in various scenarios with small training time overhead.
\end{abstract}


\section{INTRODUCTION}
Massive multiple-input multiple-output (MIMO) is one of the crucial components of beyond 5G and 6G communications \cite{Larsson2014}. Implementation of a large array of antennas can enhance the system throughput or energy efficiency through intelligent beamforming designs. However, for massive MIMO techniques to be fully effective, precise channel state information (CSI) is required \cite{Marzetta2010}. While CSI can be obtained through channel estimation techniques, estimated channels might become outdated due to system feedback delays \cite{Ramya2009} or mobility of user equipments (UEs) \cite{Papa2017}, \cite{Truong2013}. To overcome this issue, channel prediction is introduced, which utilizes the channel estimates from previous time slots without requiring additional pilot resources \cite{Kong2015}. 

Methodologies for channel prediction can be broadly classified into two categories: model-based and machine learning (ML)-based. Model-based channel predictors rely on mathematical frameworks and statistical methods to model and predict channel behaviors. For instance, in \cite{Qin2022}, the downlink channel was predicted by leveraging the partial channel reciprocity with the uplink channel and utilizing the angle-delay-Doppler structure of the channels. In \cite{Turan2024}, Gaussian mixture models were used to capture the joint distributions of channel trajectories for moving UEs, enabling prediction without the need for SNR-specific re-training. Additionally, in \cite{Wang2023}, an auto-regressive (AR) predictor was combined with genetic programming and higher-order differential equations (GPODE) to capture the time-varying nature and non-linearity of channels.

Parallel to these works, there has been an increasing interest in channel prediction techniques based on ML to tackle non-stationary and fast-varying channel environments. 
These techniques leverage a wide range of network architectures, including multi-layer perceptrons (MLPs) \cite{MLP}, sequential models like recurrent neural networks (RNNs) \cite{RNN1, RNN2, RNN3}, NeuralProphet \cite{Shehzad2022}, and transformers \cite{Hao2022}, as well as convolutional neural networks (CNNs) \cite{CNN2}, and Bayesian neural networks (BNNs) \cite{BNN}.
Specifically, for narrowband massive MIMO systems, an MLP was implemented and compared with a Kalman filter-based channel predictor in \cite{MLP}, and two types of channel predictors that utilize different RNNs, i.e., long short-term memory (LSTM) and gated recurrent unit (GRU) were compared in \cite{RNN1}. 
In \cite{Shehzad2022}, a hybrid framework that combines an RNN with a NeuralProphet was proposed. In this framework, the RNN's outputs serve as future regressors for the NeuralProphet. Additionally, \cite{Hao2022} introduced a transformer-based channel predictor that enables parallel, multiple-step prediction through an attention mechanism.
In \cite{CNN2}, a CNN was utilized to extract an auto-correlation function (ACF), which characterizes the aging feature of the channel, and combined with an AR channel predictor. Moreover, a BNN-based channel predictor was developed in \cite{BNN}, which automatically optimizes regularization hyper-parameters.

ML-based channel prediction techniques were further extended to wideband massive MIMO systems that employ the orthogonal frequency division multiplexing (OFDM) technique \cite{OFDM1, OFDM2, OFDM3}. The channel predictions for MIMO-OFDM systems were performed in different channel domains, such as an array-frequency domain or angle-delay domain. In \cite{OFDM1}, array-frequency domain channels, which are the common form of matrix channels for MIMO-OFDM systems, were transformed into angle-delay domain channels for higher multipath angle and delay resolutions, and a spatio-temporal auto-regressive (ST-AR) predictor was proposed to predict significant elements of the sparse angle-delay domain channels using a complex-valued neural network (CVNN). Meanwhile,  in \cite{OFDM2}, an attention mechanism with differencing operation was implemented to exploit the spatio-temporal correlations of array-frequency domain channels via ConvLSTM, which is a combination of convolutional layers and LSTM networks. In \cite{OFDM3}, array-frequency domain channels were interpreted as multiple array domain channels, and an MLP trained with array domain channels from a certain subcarrier was used to predict the array domain channels from all subcarriers by exploiting the high correlations between different array domain channels.

While ML-based channel predictors have shown promise, those trained offline may suffer when they encounter channel environments that differ significantly from what they were trained on. This is due to the fact that the neural network will be trained to handle the scenarios described by the training dataset. Hence, when the neural network encounters vastly different channel behaviors such as the channels after a long time period \cite{Wang2023} or the channels from different UEs \cite{Kim2023}, its performance may worsen considerably. 

To tackle these challenges, we propose an online re-training framework for ML-based channel prediction. This framework follows a cyclical, open-loop process, alternating between training and prediction phases. In the training phase, new training data that reflect recent environmental changes are collected, and then the neural network is re-trained from scratch using these data. In the prediction phase, the channels are forecasted using the newly trained neural network. This process allows the system to update ML-based channel predictors to reflect new UEs or evolving channel conditions, thereby improving the predictive accuracy.

Within the online re-training framework, our objective is to design an ML-based channel prediction technique that efficiently utilizes a limited amount of training data. This design criterion is motivated by the need to reduce the time overhead during the training phase to develop practical channel predictors. The training time overhead encompasses both data collection and computation time for network training. First, in tasks involving the prediction of temporal data, the data collection time is directly related to the size of the training dataset, leading to a smaller dataset when reducing the data collection time. However, a sufficient number of training data is required to train neural networks properly and to prevent over-fitting as indicated in \cite{Training}. Consequently, it is essential to design ML-based channel predictors that not only mitigate the risk of performance degradation due to insufficient data but also reduce the data collection time. Second, the computation time for network training is influenced by multiple factors such as network architectures, hyper-parameters, and optimization algorithms. Moreover, computation time varies significantly with the hardware specifications used for training, which are not inherently related to the structure of ML-based channel predictors themselves. Considering these factors and assuming that advanced hardware, e.g., neural processing unit (NPU), can accelerate the computation time, we focus on developing an ML-based channel predictor that addresses the challenges of data scarcity with the need to reduce data collection time.

While conventional ML techniques such as data augmentation and meta-learning are commonly employed to address the scarcity of a training dataset, we propose a unique approach that diverges from these techniques. Data augmentation, which can expand the training dataset by generating synthetic data \cite{DATAAUG1, DATAAUG2}, needs careful implementation to ensure that the synthetic data closely mirror the true data distribution; otherwise, there is a risk of over-fitting to artificially created data \cite{DATAAUG3}. Meta-learning can be also advantageous when the number of training data is limited \cite{Timothy2022}. In particular, channel prediction via meta-learning in \cite{Kim2023} enables quick adaptation of network to a new environment using minimal data. However, meta-learning requires additional data collection and training during the meta-training stage. Therefore, we propose a novel approach tailored to the online re-training framework, which neither artificially creates data nor requires additional training stage as in meta-learning.

In this paper, we propose an ML-based channel prediction technique called aggregated learning (AL) approach for wideband massive MIMO systems that allows sufficient amount of training data with reduced time for data collection. The main feature of the AL approach is the pre-processing of training data. 
We demonstrate two distinct variants of the AL approach called AL in the array domain (AL-AD) and AL in the frequency domain (AL-FD) by interpreting an array-frequency domain channel of MIMO-OFDM systems as multiple array domain channels or multiple frequency domain channels, respectively, which are the channel domains for data pre-processing. In AL-AD, the training data represented in the array-frequency domain are split into multiple sub-data represented in the array domain, and the multiple sub-data are aggregated\footnote{In this paper, when data are aggregated, it means that they have been gathered or combined from multiple sources to form a unified dataset.} into a new training dataset. And then, a neural network is trained with the new training dataset, which contains the aggregated characteristics of the array domain channels from multiple subcarriers.
Finally, the array-frequency domain channel for the MIMO-OFDM system is reconstructed from the predicted array domain channels from every subcarrier. Similarly, in AL-FD, the training data represented in the array-frequency domain are pre-processed into multiple sub-data represented in the frequency domain, and the predicted frequency domain channels from all antennas are reconstructed into the array-frequency domain channel.
While both AL-AD and AL-FD utilize the same training data from the array-frequency domain, the prediction performance varies due to the distinct channel domain for data pre-processing. Based on numerical results, it appears that both AL-AD and AL-FD can decrease the time overhead during the training phase while AL-FD outperforms other benchmark channel predictors including AL-AD owing to its unique channel form. The contributions of this paper are summarized as follows:

\begin{itemize}
	\item We introduce an online re-training framework designed to improve the performance of ML-based channel predictors under dynamic channel conditions. This framework operates through a cyclical process, alternately re-training the neural network from scratch with new training data and predicting channel. This approach ensures continuous updates to reflect new UEs or evolving channel conditions, thereby improving predictive accuracy. With an emphasis on practical deployment, the framework is considered in scenarios where data collection time is limited.
	\item We propose the AL approach for ML-based channel prediction in wideband massive MIMO systems, aimed at reducing data collection time by pre-processing original training data to increase the volume of training dataset. The AL approach features two variants, AL-AD and AL-FD. While both variants reduce training time overhead, AL-FD demonstrates enhanced prediction performance, leveraging its domain-specific data pre-processing.
	\item We analyze three types of correlation properties of array-frequency domain channels to understand how the AL approach exploits spatio-temporal correlations. Initially, we explore the correlation among multiple split sub-channels to assess the diversity impact on the training dataset. Subsequently, we examine the partial spatio-temporal correlations of each type of sub-channel by first analyzing the correlation among elements within the sub-channel to evaluate how the elements are spatially correlated, and then investigate temporal correlation to assess how the sub-channels evolves over time. These analyses provide a key for comprehending the prediction performance enhancements provided by the AL approach.	
\end{itemize}

The rest of paper is organized as follows. A system model and channel estimation are described in Section \ref{System}. In Section \ref{Online}, we explain the motivation behind the online re-training framework for ML-based channel prediction and emphasize the necessity to minimize training time overhead. We propose the AL approach that pre-processes training data in Section \ref{Proposed}. In Section \ref{Correlation}, we examine three types of correlation properties within the array-frequency domain channel, which can be used to analyze the performance of proposed AL approach. In Section \ref{Numerical}, we present numerical results of channel prediction techniques based on ML. Finally, we conclude the paper in Section \ref{Conclusion}.

\textbf{Notation:} 
Column vectors and matrices are represented using lowercase and uppercase bold letters, respectively. 
The zero vector of size $m \times 1$ is denoted as $\mathbf{0}_{m}$, while $\mathbf{I}_m$ stands for the $m \times m$ identity matrix.
The multivariate complex normal distribution, characterized by a mean vector $\boldsymbol{\mu}$ and a covariance matrix $\boldsymbol{\Sigma}$, is denoted as $\mathcal{CN}(\boldsymbol{\mu}, \boldsymbol{\Sigma})$.
Expectation and trace are symbolized by $\mathbb{E} \left[ \cdot \right]$ and $\text{tr} \left[ \cdot \right]$, respectively. The magnitude of a vector is indicated as $\lVert \cdot \rVert$. The set of all $m \times n$ complex matrices is denoted as $\mathbb{C}^{m \times n}$.
For a complex number $a$, $\text{Re}\left(a \right)$ and $\text{Im}\left(a \right)$ denote its real and imaginary components, respectively. $\mathcal{O}$ is employed for Big-O notation, while $\text{vec}(\cdot)$ denotes a function that converts a matrix into a vector. The $i$-th row and the $j$-th column of a matrix $\bA$ are denoted as $\bA[i, :]$ and $\bA[:, j]$, respectively, and the $(i, j)$-th element of the matrix $\bA$ is denoted as $\bA[i, j]$.

\section{SYSTEM MODEL AND CHANNEL ESTIMATION}\label{System}
In this section, we introduce a system model for a wideband massive MIMO and describe a channel estimation technique, where the estimated channels are utilized for ML-based channel prediction.

\subsection{SYSTEM MODEL}

\begin{figure}[t]
	\centering
	\includegraphics[width=1.0\columnwidth]{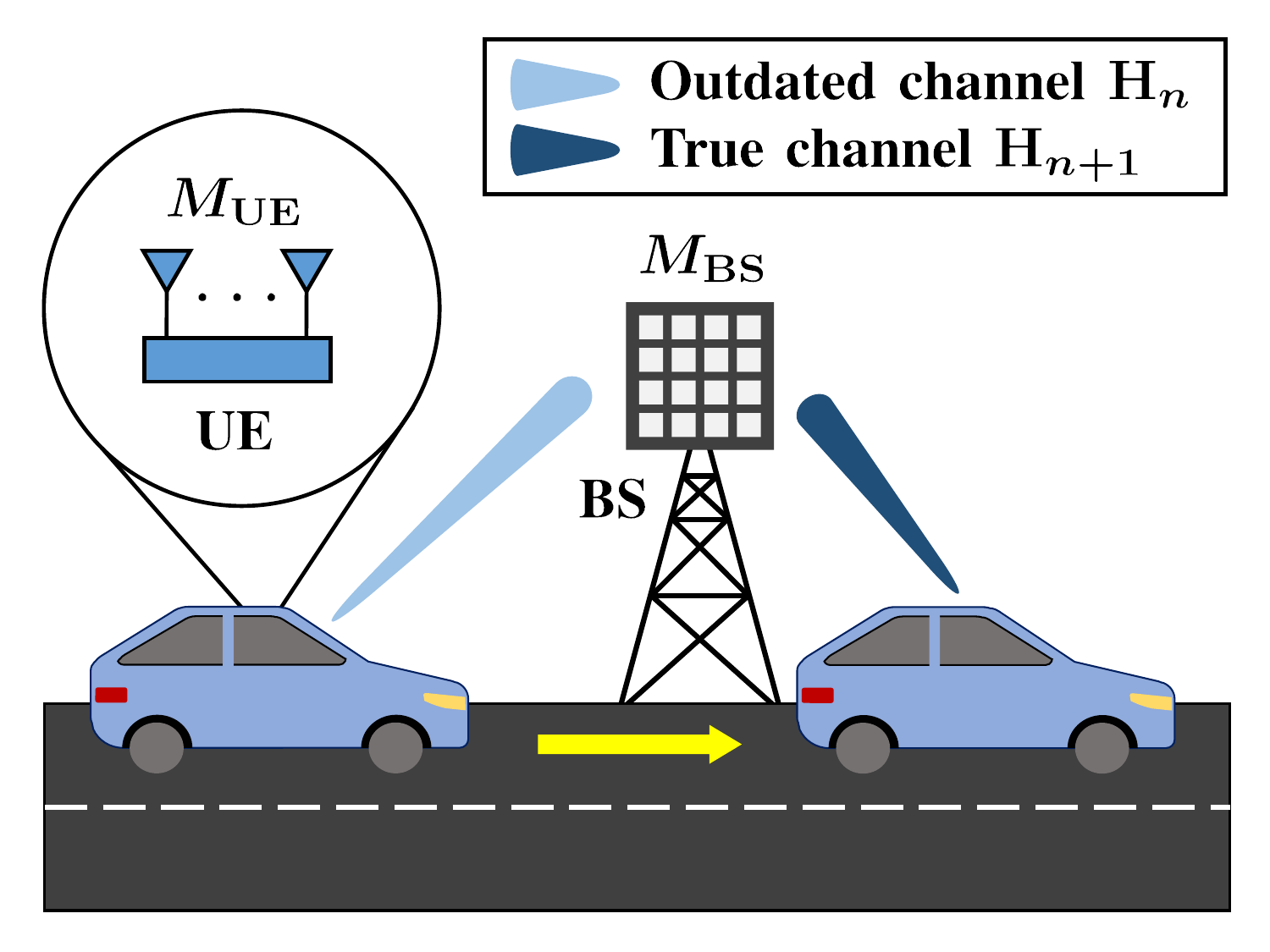}
	\caption{A massive MIMO system consisting of a BS with $M_\text{BS}$ antennas and a UE with $M_\text{UE}$ antennas, where the estimated channels can be outdated due to the mobility of the UE.}\label{fig: System}
\end{figure}

As the base station (BS) has the ability to estimate and predict the channel for each UE separately, we consider an uplink wideband massive MIMO system where a UE with $M_\text{UE}$ antennas communicates with a BS with $M_\text{BS}$ antennas as described in Fig. \ref{fig: System}.
The total number of antenna pairs in the system is denoted as $M=M_\text{BS}M_\text{UE}$.
The OFDM technique is utilized to resolve inter-symbol-interference (ISI) problem in the wideband system, which converts the frequency-selective fading channel into $L$ flat fading channels with separate subcarriers. In the MIMO-OFDM system, the uplink received signal at the $n$-th time slot for the $\ell$-th subcarrier is given as 
\begin{align}\label{eq: eq1}
	\by^{\ell}_n= \bH^{\ell}_{n} \bx^{\ell}_{n} + \bw^{\ell}_n,	
\end{align}
where $\bH^{\ell}_{n} \in \mathbb{C}^{M_\text{BS} \times M_\text{UE}}$ is the MIMO channel, $\bx^{\ell}_{n} \in \mathbb{C}^{M_\text{UE} \times 1}$ is the transmit signal, and $\bw^{\ell}_n \sim \mathcal{CN}\left(\mathbf{0}_{M_\text{BS}}, \sigma^2 \bI_{M_\text{BS}}  \right)$ is complex additive white Gaussian noise (AWGN) with noise variance $\sigma^2$.

\subsection{CHANNEL ESTIMATION}
In the channel estimation phase, the received length $\tau$ pilot signal at the $n$-th time slot for the $\ell$-th subcarrier is given as
\begin{align}\label{eq: eq2}
	\bY^{\ell}_n = \sqrt{\rho} \bH^{\ell}_n  {\boldsymbol{\Phi}^{\ell}_n}^{\mathrm{T}} + \bW^{\ell}_n,
\end{align}
where $\rho$ is the pilot power, $\boldsymbol{\Phi}^{\ell}_n \in \mathbb{C}^{\tau \times M_\text{UE}}$ is the pilot matrix assuming the column-wise orthogonality, i.e., ${\boldsymbol{\Phi}^{\ell}_n}^\mathrm{H} {\boldsymbol{\Phi}^{\ell}_n}=\tau \bI_{M_\text{UE}}$, and $\bW^{\ell}_{n} \in \mathbb{C}^{M_\text{BS} \times \tau}$ is the complex AWGN. 
Using the vectorization function, the received pilot signal is transformed into 
\begin{align}\label{eq: eq3}
	{\bar{\by}}^\ell_n & = \text{vec}\left( \bY^{\ell}_n \right)  \nonumber \\ 
	&= \boldsymbol{\Psi}^\ell_n {\bh}^\ell_n + {\bw}^\ell_n,
\end{align}
where $\boldsymbol{\Psi}^\ell_n=\sqrt{\rho}\left({\boldsymbol{\Phi}^{\ell}_n} \otimes \bI_{M_\text{BS}} \right) \in \mathbb{C}^{\tau M_\text{BS} \times M}$, ${{\bh}}^\ell_n = \text{vec}\left( \bH^{\ell}_n \right) \in \mathbb{C}^{M \times 1}$ is the vectorized MIMO channel, and ${{\bw}}^\ell_n = \text{vec}\left( \bW^{\ell}_n \right) \in \mathbb{C}^{\tau M_\text{BS} \times 1}$.
Using the least square (LS) method with the condition $\tau \ge M_\text{UE}$, the estimated vectorized MIMO channel at the $n$-th time slot for the $\ell$-th subcarrier is given as
\begin{align}\label{eq: eq4}
	\bg_n^{\ell} = {{\bh}}^\ell_n + \tilde{\bw}^\ell_n,
\end{align}
where $\tilde{\bw}^\ell_n = \left( {\boldsymbol{\Psi}^\ell_n}^\mathrm{H} \boldsymbol{\Psi}^\ell_n \right)^{-1} {\boldsymbol{\Psi}^\ell_n}^\mathrm{H} {\bw}^\ell_n$ is the channel estimation error, which is distributed as $\tilde{\bw}^\ell_n \sim \mathcal{CN}\left(\mathbf{0}_M, \frac{\sigma^2}{\rho \tau}\bI_M  \right)$.

While the channel estimation is performed for each subcarrier, the estimated channel for the entire system can be collected into a matrix $\bG_n \in \mathbb{C}^{M \times L}$, where the $\ell$-th column corresponds to the estimated vectorized MIMO channel for the $\ell$-th subcarrier. In this paper, we describe this matrix as the array-frequency domain channel, which captures the entire channel coefficients of all antenna and subcarrier pairs of the MIMO-OFDM system.

In our scenario of interest, channel estimation is initially conducted at the $n$-th time slot, estimating $\mathbf{H}_n \in \mathbb{C}^{M \times L}$, which gives the estimated channel $\mathbf{G}_n$. However, due to rapid changes in channel conditions, the estimated channel $\mathbf{G}_n$ becomes outdated as the true channel is now $\mathbf{H}_{n+1} \in \mathbb{C}^{M \times L}$, which is the channel at the next time slot, as described in Fig. \ref{fig: System}. To address this issue, we develop an ML-based channel prediction approach that utilizes the past estimates\footnote{Note that the estimated channels are used instead of true channels since obtaining true channel values is not possible in practice.} $\left\{\bG_i\right\}_{i \leq n}$ to predict the true channel $\bH_{n+1}$, where the predicted channel at the $n$-th time slot is denoted as $\hat{\bH}_{n+1}\in \mathbb{C}^{M \times L}$.

\section{ONLINE RE-TRAINING FRAMEWORK FOR ML-BASED CHANNEL PREDICTION}\label{Online}

In this section, we begin by outlining the motivation for implementing an online re-training framework for ML-based channel prediction. Next, we describe the framework in detail, emphasizing the design and functionality. Lastly, we highlight the importance of minimizing the training time overhead in the framework.

\subsection{MOTIVATION}
While many studies have concentrated on offline training \cite{MLP, RNN1, RNN2, RNN3, CNN2, BNN, Shehzad2022, Hao2022, OFDM1, OFDM2, OFDM3, Offline1, Offline2, Kim2023}, which show promise in stable, controlled environments, they often overlook the dynamic and evolving nature of real-world wireless systems. In wireless systems, the BS consistently encounters UEs, which are continuously in motion, in diverse environments. Consequently, a predictor that has been pre-trained offline using data from a particular UE may not retain the prediction accuracy when applied to new UEs or sustain its performance over time due to inevitable fluctuation in channel statistics. The online re-training framework is designed to train the ML-based channel predictor from scratch in real-time, considering the frequent introduction of new UEs to the system or the temporal evolution of channel statistics. Note that our proposed framework does not suffer from issues such as catastrophic forgetting, negative transfer, or domain shift, which are critical challenges in continual learning, transfer learning, and meta-learning, as our primary objective is to train the network from scratch for the current wireless environment without retaining information from past environments.

\begin{figure}
	\centering
	\includegraphics[width=1.0\columnwidth]{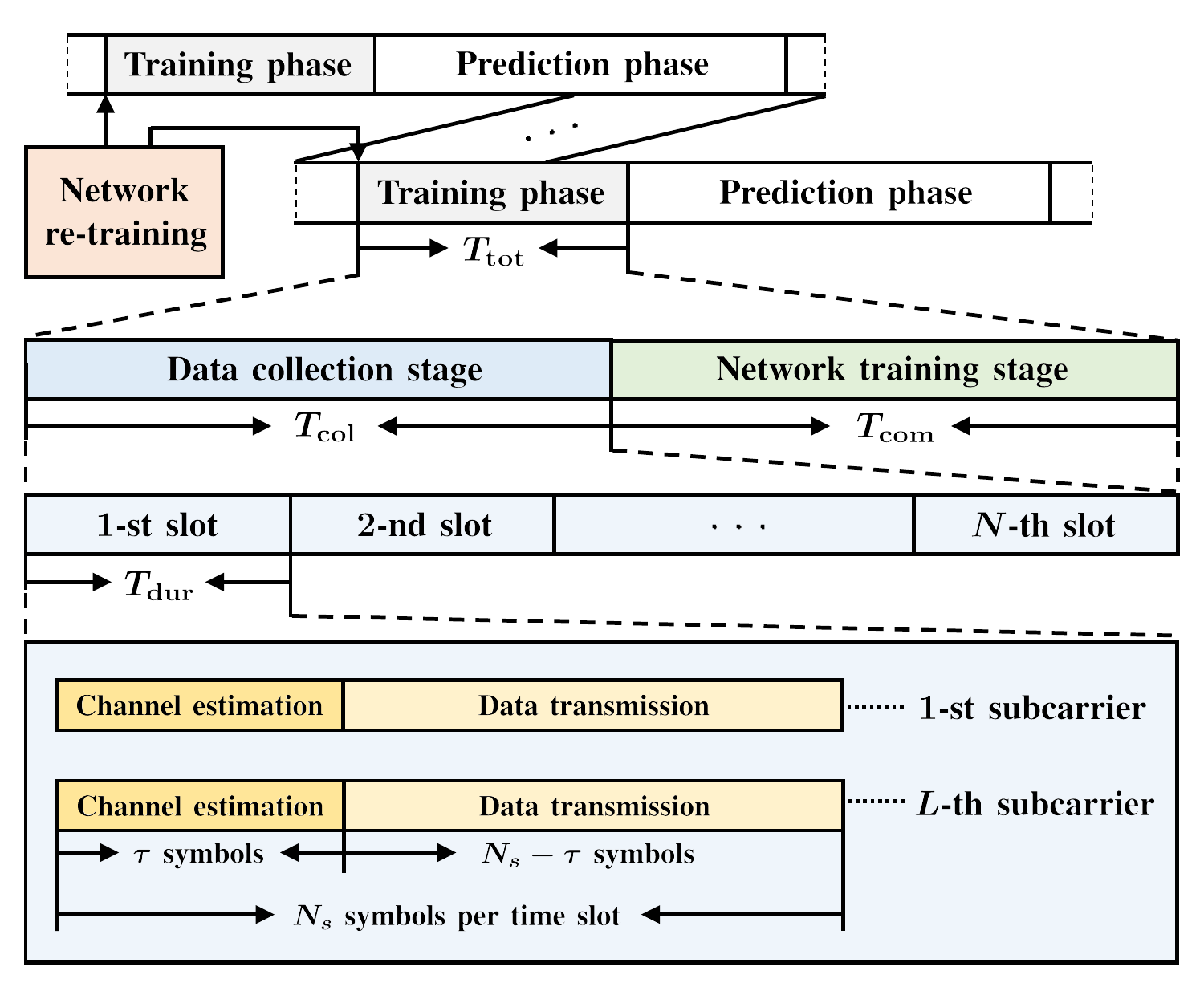}
	\caption{Overall process of online re-training framework.}\label{fig: Process}
\end{figure}

\subsection{ONLINE RE-TRAINING FRAMEWORK}
The online re-training framework for ML-based channel prediction is characterized by a cyclical process of training and prediction phases, as illustrated in Fig. \ref{fig: Process}. Arising from the motivation discussed earlier, the framework initiates the network re-training by entering the training phase to update the neural network with new training dataset, ensuring the predictor remains current and effective. Within the training phase, there are two stages, the data collection stage and the network training stage. Initially, new training data are gathered based on recent channel estimates, reflecting the latest environmental changes. Following the data collection, the neural network is trained from scratch with the updated dataset to refine its predictive capabilities.

After completing the training phase, the framework transitions into the prediction phase, where the updated network is employed to predict the channels. This iterative approach, alternating between training and prediction, ensures the predictor is continuously updated to reflect new UEs and changing channel statistics, thereby maintaining high prediction accuracy over time. It is important to note that the duration of the training phase, which is regarded as the training time overhead, plays a critical role in the design of practical ML-based channel prediction techniques, since excessive overhead can disrupt the BS from acquiring accurate channel values promptly, resulting in the performance degradation of massive MIMO systems \cite{Marzetta2010}. Thus, reducing the training time overhead is essential for sustaining the system performance.

\subsection{TRAINING TIME OVERHEAD}
The time overhead of the training phase is modeled as
\begin{align}
	T_\text{tot} = T_\text{col} + T_\text{com},  
\end{align}
where $T_\text{col}$ is the time for the data collection, and $T_\text{com}$ is the computation time consumed during the neural network training. The time for data collection is defined as $T_\text{col}=T_\text{dur} \cdot N$, where $T_\text{dur}$ is the duration of each time slot, and $N$ is the number of time slots allocated for data collection. Within each time slot of $N_s$ symbols, $\tau$ symbols are dedicated to the channel estimation for every subcarrier.

While minimizing both the data collection and computation time is crucial for the efficiency of the online re-training framework, addressing the computation time is a non-trivial problem due to influences from the choices of neural networks, hyper-parameters, optimization algorithms, and even hardware specifications. Hence, we focus toward reducing the data collection time, which provides a more straightforward approach by limiting the number of time slots for data collection. 

However, training a neural network with a limited dataset, such as a dataset collected from fewer than a hundred time slots, presents a significant challenge due to the scarcity of training data \cite{Training}, especially when conventional ML-based channel predictors for MIMO-OFDM systems typically require more than hundreds to thousands of time slots for the data collection \cite{OFDM2}. Therefore, our goal is to develop an ML-based channel prediction approach that overcomes the issue of insufficient training data resulting from a constrained data collection time.

\section{AL APPROACH}\label{Proposed}
In this paper, we focus on predicting channels within one cycle of the online re-training framework in Fig. \ref{fig: Process}, where the proposed AL approach can be applied to each cycle of online re-training.
The AL approach introduces a method for pre-processing the training data to ensure the neural network is supplied with an up-to-date and a sufficient quantity of training data.

\subsection{TRAINING DATA}
To exploit the temporal correlation of the channel, the training data is constructed from a sequence of estimated array-frequency domain channels and denoted as
\begin{align}
	\left(X_n, Y_n \right)=\left( \left\{{{\bG}}_{k}\right\}_{k=n-I+1}^n, {{\bG}}_{n+1} \right), \ n \in \mathcal{N}_\text{tr}, 
\end{align}
where $X_n=\left\{{{\bG}}_{k}\right\}_{k=n-I+1}^n$ is the feature, $Y_n={{\bG}}_{n+1}$ is the label focusing on one-step prediction,\footnote{For multiple-step ahead prediction, the structure of the training data $\left(X_n, Y_n \right)$ can be adjusted by changing the label from $Y_n={{\bG}}_{n+1}$ to $Y_n=\left\{{\bG_k}\right\}_{k=n+1}^{n+p}$, where $p$ denotes the prediction order.} $I$ is the input order that depends on the mobility of the UE \cite{MLP}, and $\mathcal{N}_\text{tr}$ is the set of consecutive time slots for the training phase. Then, the training dataset collected during $N$ time slots is denoted as 
\begin{align}
	\mathcal{D} = \left\{\left(X_n, Y_n \right) | n \in \mathcal{N}_\text{tr} \right\},
\end{align}
where $\left\vert \mathcal{D} \right\vert=\left\vert \mathcal{N}_\text{tr} \right\vert=N-I$. 
Given the necessity to keep $N$ relatively small to minimize the data collection time within the online re-training context, directly utilizing the dataset $\mathcal{D}$ might not provide an adequate number of training data for effective network training. Therefore, in the AL approach, the training data are processed before being introduced to the neural network.

\subsection{SUB-CHANNELS}
In the AL approach, the array-frequency domain channel \(\bH_n\) is decomposed into multiple sub-channels, which are categorized into two types: 1) the array domain channels and 2) the frequency domain channels. The array domain channels are defined by the columns of $\bH_n$, where the $\ell$-th column, $\bH_n[:, \ell]$, is the array domain channel that the $\ell$-th subcarrier sees. Similarly, the second type of sub-channels are the frequency domain channels, where the $m$-th row of $\bH_n$ is the frequency domain channel observed from the $m$-th antenna, i.e, $\bH_n[m, :]$. Please note that the array domain channels are the common representation of channels in MIMO-OFDM systems while the frequency domain channels offer a novel perspective of MIMO-OFDM channels. 
For the simplicity in discussion of the AL approach, we denote both types of sub-channels within the array-frequency domain channel $\bH_n$ as $\bH_n^i[:] \in \mathbb{C}^{K_1 \times 1}$ for $i=1, \cdots, K_2$, where $K_1$ is the size of the sub-channel, and $K_2$ is the number of sub-channels.
Hence, the sub-channels can be either $\bH_n^i[:] = \bH_n[:, i]$ for the array domain with $K_1=M$ and $K_2=L$ or $\bH_n^i[:] = \bH_n[i, :]^\mathrm{T}$ for the frequency domain with $K_1=L$ and $K_2=M$.  

\subsection{DATA PRE-PROCESSING}
\begin{figure}[t]
	\subfloat[Data pre-processing of the AL approach. ]{{\includegraphics[width=0.485\textwidth ]{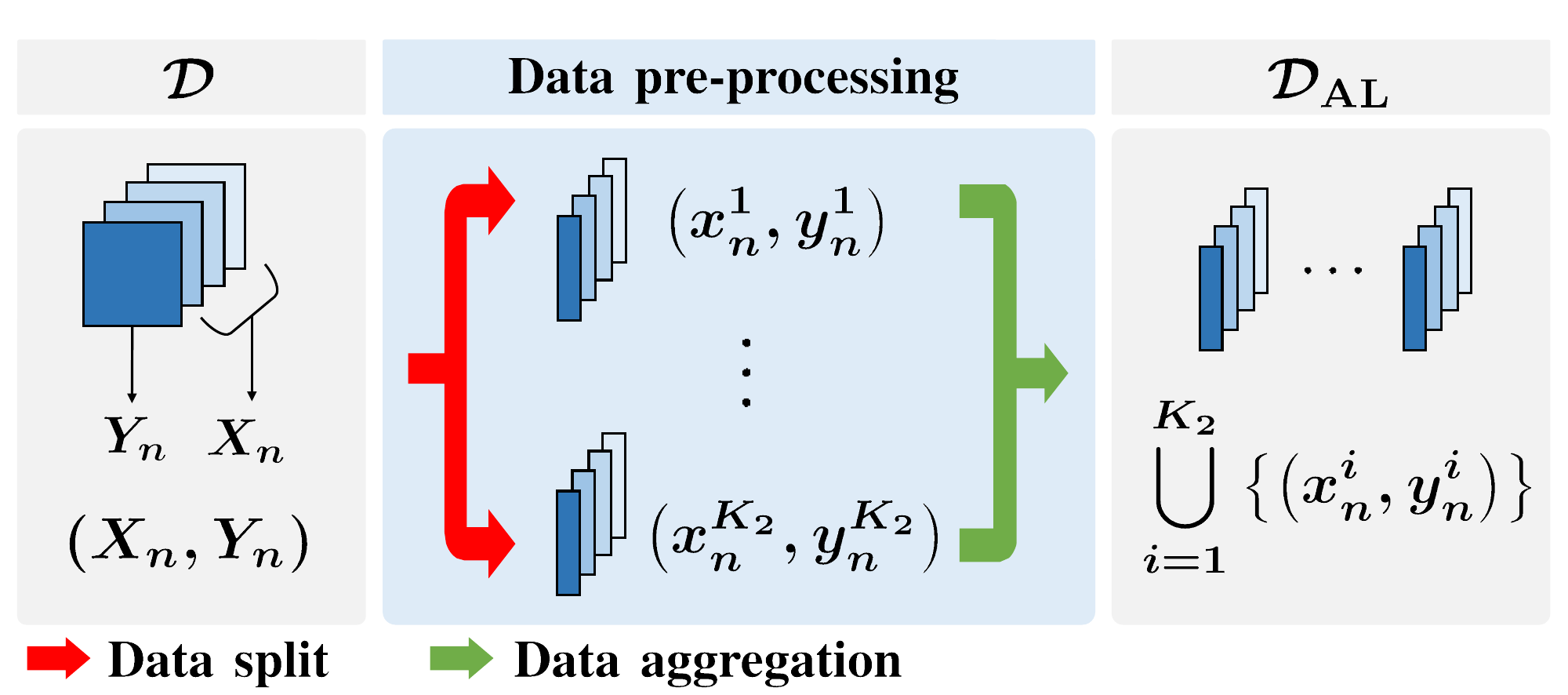} }} 		
	\centering			
	\vfill
	\subfloat[Network training of the AL approach.]{{\includegraphics[width=0.485\textwidth ]{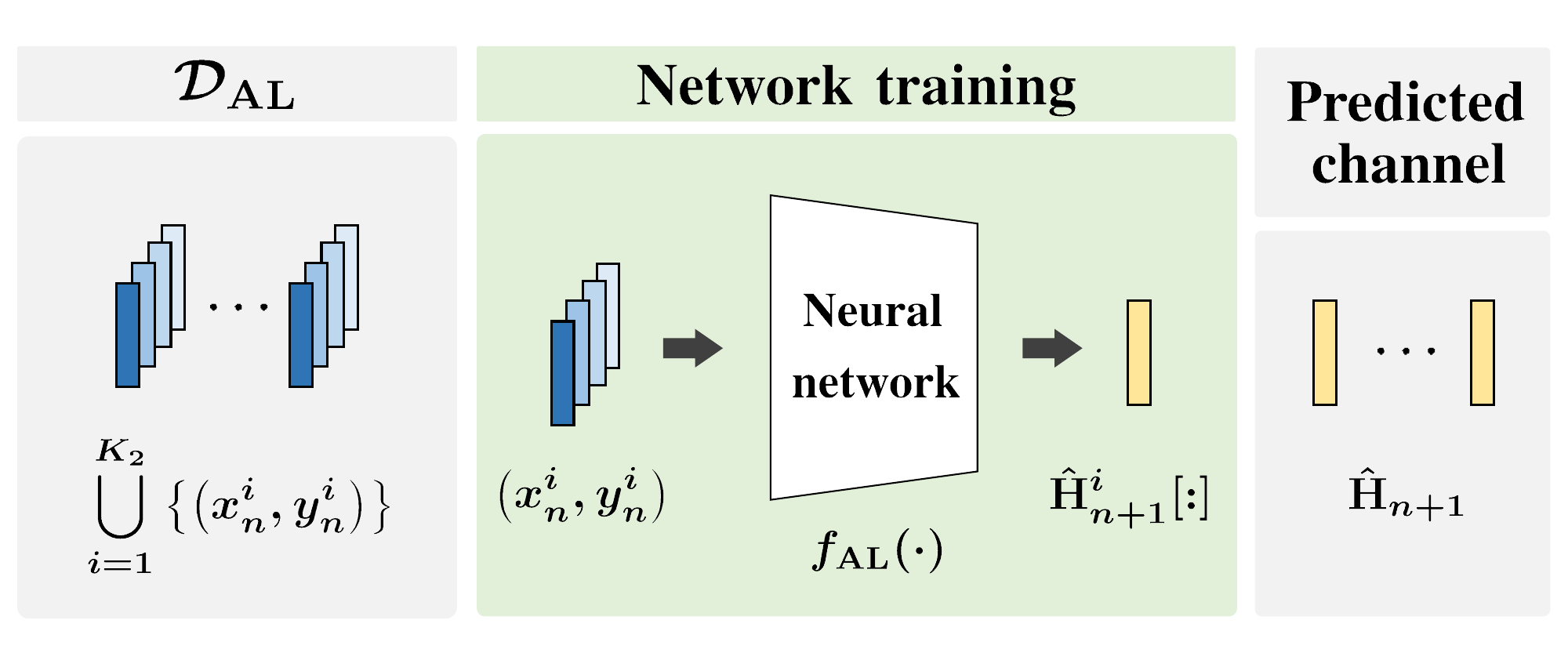} }}
	\centering		
	\caption{Overall process of the AL approach.} 		
	\label{fig: AL}	
\end{figure}

The insight that the array-frequency domain channel consists of $K_2$ sub-channels, such as $L$ array domain channels or $M$ frequency domain channels, provides an alternative domain for collecting the training data with the limited resource in the time domain. By decomposing the array-frequency domain channel into multiple sub-channels, the training data  $\left(X_n, Y_n\right)$ can be split into multiple sub-data as described in Fig. \ref{fig: AL} (a). The $i$-th sub-data is denoted as
\begin{align}
	\left(x_n^i, y_n^i \right)=\left( \left\{{{\bG}}_k^i[:] \right\}_{k=n-I+1}^n, {{\bG}}_{n+1}^i[:] \right), \ n \in \mathcal{N}_\text{tr}, 
\end{align}
where ${{\bG}}_{n}^i[:]$ is the estimated $i$-th sub-channel, and $x_n^i$ and $y_n^i$ are the feature and label represented with the $i$-th sub-channel, respectively.
Next, $K_2$ sub-data are aggregated into a new training dataset given as
\begin{align}
	\mathcal{D}_\text{AL} = \bigcup_{i=1}^{K_2} \left\{ \left(x_n^i, y_n^i \right) | n \in \mathcal{N}_\text{tr} \right\}, 
\end{align}
where $\left\vert \mathcal{D}_\text{AL} \right\vert=K_2(N-I)$ is the total number of training data in the AL approach. 
Note that the total amount of training data in the AL approach will increase in proportion to $K_2$ compared to the number of training data in $\mathcal{D}$. 

\subsection{NEURAL NETWORK TRAINING}
After the data pre-processing, the neural network undergoes training with the new training dataset as illustrated in Fig. \ref{fig: AL} (b). It is important to note that a specific network architecture for the neural network in the AL approach is not explicitly defined. This is because the core focus of the AL approach is on the novelty of the data pre-processing rather than on any particular type of network architecture. The compatibility of various network architectures, such as MLP, RNN, LSTM, and transformer, within the AL approach is demonstrated in Section \ref{Numerical}.

Simple reshaping of features and labels to match the input and output dimensions of different network architectures ensures adaptability. For example, in the case of an MLP, each sub-data $\left(x_n^i, y_n^i\right)$ is split into real and imaginary parts and then vectorized. This reshapes the feature and label into vectors with dimensions $2IK_1 \times 1$ and $2K_1 \times 1$, respectively, aligning with the MLP. Similarly, for RNN, LSTM, and transformer, the feature and label are reshaped into a matrix and a vector of size $I \times 2K_1$ and $1 \times 2K_1$, respectively.
The real output values from a neural network are then recombined into complex values, which correspond to the predicted channel.
For the optimizer, an adaptive moment estimation (ADAM) in \cite{ADAM} is employed, and the loss function is the mean square error (MSE) between the estimated and predicted channels. The loss function for the AL approach is given as
\begin{align}
	\text{Loss}_\text{AL}=\frac{1}{|\mathcal{N}_\text{tr}|}\frac{1}{K_2}\sum_{n \in \mathcal{N}_\text{tr}} \sum_{i=1}^{K_2} \left\Vert \bG_{n+1}^i[:] - \hat{\bH}_{n+1}^i[:] \right\Vert_2^2,
\end{align}
where $\hat{\bH}_{n+1}^i[:]$ is the predicted $i$-th sub-channel. Although the loss function compares the predicted channel $\hat{\bH}_{n+1}^i[:]$ with the estimated channel $\bG_{n+1}^i[:]$ from the $(n+1)$-th time slot, this comparison occurs within the training phase. The estimated channels from the $(n+1)$-th time slot are not the future values in the sense of being unknown or unavailable at the time of training, but they have already been collected and are used to construct the training data. In this context, the estimated channel from the $(n+1)$-th time slot serves as the label for the training data, facilitating the learning of the temporal evolution of the channel.

\subsection{CHANNEL PREDICTION}
After completing the training phase, the trained network is deployed to predict channels. The prediction for the $i$-th sub-channel is formulated as 
\begin{align}
	\hat{\bH}_{n+1}^i[:] = f_{\text{AL}}\left(\left\{\bG_k^i[:] \right\}_{k=n-I+1}^n \right), \ n \in \mathcal{N}_\text{pr},
\end{align}
where $f_{\text{AL}}(\cdot)$ is the neural network that is trained to predict every sub-channels. The input $\left\{\bG_k^i[:] \right\}_{k=n-I+1}^n$ is the past estimates of the $i$-th sub-channel, and $\mathcal{N}_\text{pr}$ is the set of time slots during the prediction phase, which is disjoint from the set of time slots for the training phase. The array-frequency domain channel is reconstructed based on the predictions of $K_2$ sub-channels. Based on the two types of sub-channels, the AL approach provides two variants called AL-AD and AL-FD, which perform the AL approach in the array domain and frequency domain, respectively. 
While both AL-AD and AL-FD pre-process the same training data, the prediction performance will differ due to the distinct characteristics of the array domain channels and frequency domain channels that will become clear in Sections \ref{Correlation} and \ref{Numerical}.

\textit{Remark 1: }The unique aspect of the AL approach, as opposed to conventional ML-based channel predictors for MIMO-OFDM systems, resides in the unique understanding of the channel. For instance, an ML-based channel predictor in \cite{OFDM2} predicts the array-frequency domain channel by leveraging the entire spatio-temporal correlations (both array and frequency domains jointly) by utilizing ConvLSTM architecture for the neural network. 
In contrast, the AL approach focuses on the sub-channels, rather than the array-frequency domain channel. Specifically, the AL approach predicts the sub-channels by leveraging the aggregated information across multiple sub-channels. It exploits the partial spatio-temporal correlations in either the array or frequency domain independently. This approach ensures not only the sufficient amount of training data within the limited time slots for data collection, but also offers two different channel domain options for prediction, i.e., AL-AD and AL-FD. This dual-option prediction provides more flexibility in MIMO-OFDM systems, where partial spatio-temporal correlations can vary according to specific environments.

\section{CORRELATION PROPERTIES}\label{Correlation}
In this section, we explore three types of correlation properties to understand how AL-AD and AL-FD leverage the inherent characteristics of the array-frequency domain channel. First, we examine the correlation among the split $K_2$ sub-channels, which relates to the diversity of the training dataset in the AL approach. Next, we investigate the partial spatio-temporal correlations of two types of sub-channels, by first analyzing the element-wise correlation among $K_1$ components within each sub-channel, and then exploring the temporal correlation to understand their evolution over time. These detailed examinations are crucial for understanding the prediction performance of the AL approach, which will be discussed in Section \ref{Numerical}.

\subsection{TYPE-I CORRELATION}\label{Correlation-A}

\begin{figure}[t]
	\subfloat[Correlation between $L$ array domain channels.]{{\includegraphics[width=0.5\textwidth ]{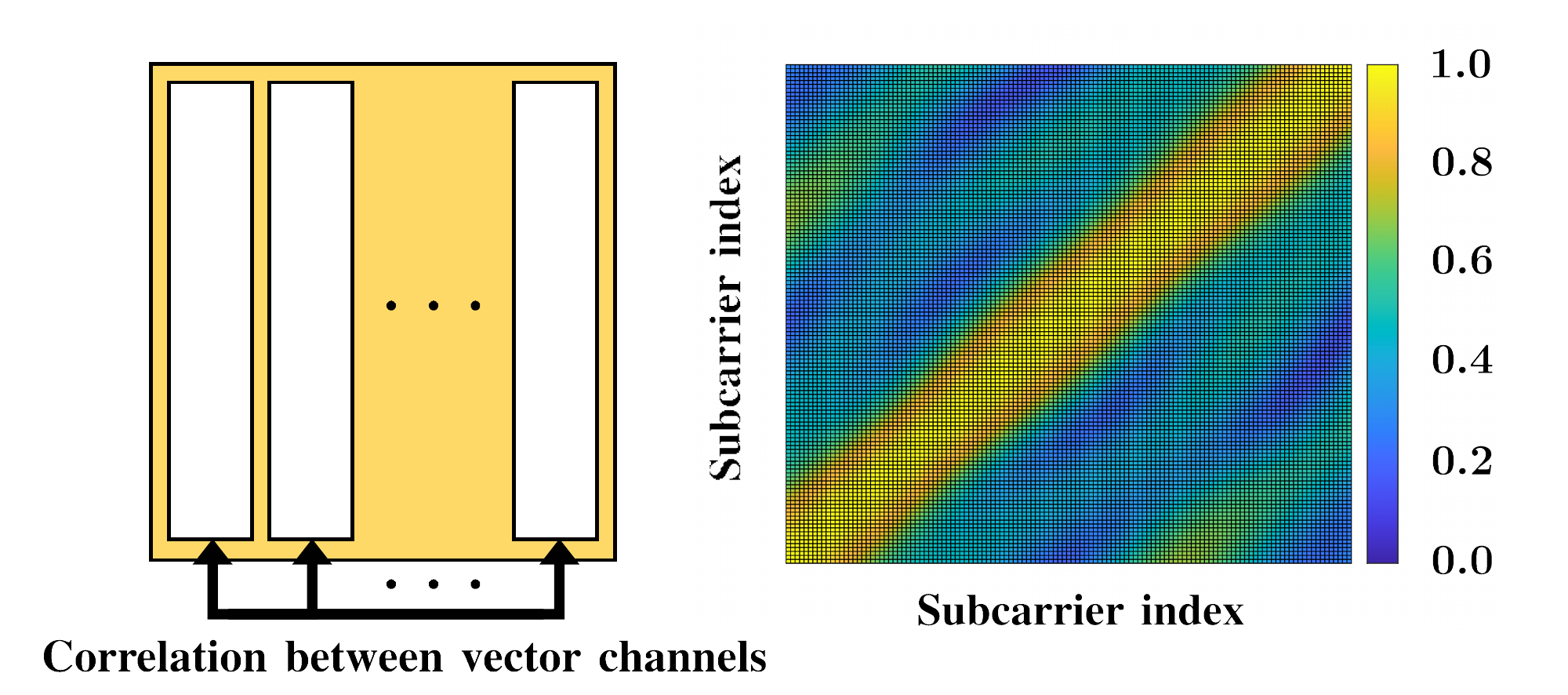} }}		
	\centering			
	\vfill
	\subfloat[Correlation between $M$ frequency domain channels.]{{\includegraphics[width=0.5\textwidth ]{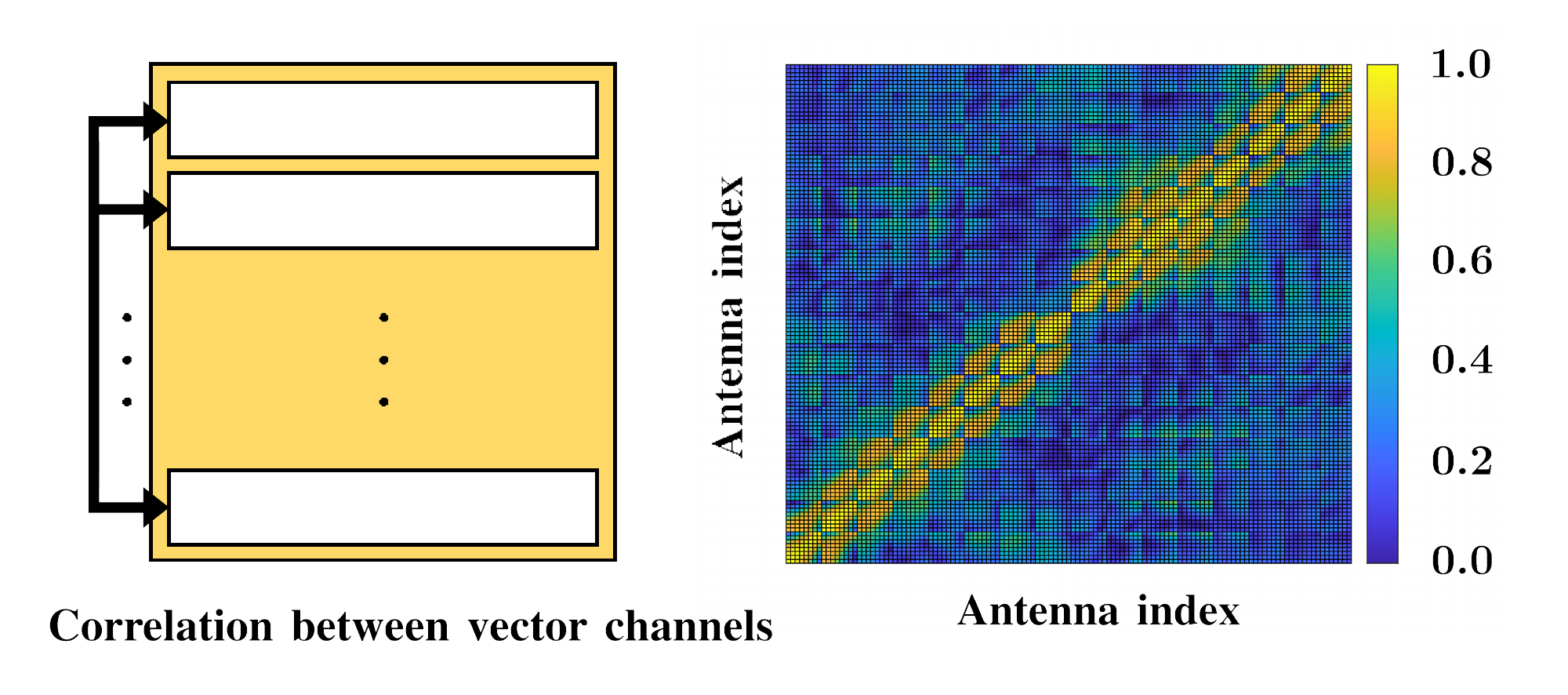} }}
	\centering		
	\caption{Type-I correlation of the array domain channels and frequency domain channels.}		
	\label{fig: Type1}		
\end{figure}

The first type of correlation property, denoted as Type-I correlation, measures the correlation between two different vector channels. This property is particularly important for diversity of the training dataset in AL-AD and AL-FD, as both channel predictors use training datasets generated by aggregating the multiple sub-data from array domain channels or from frequency domain channels. When multiple sub-data are aggregated from highly correlated channels, it may result in low diversity in the training dataset, which induces inefficient use of ML \cite{Diversity, Redundancy}. Therefore, as described in Fig. \ref{fig: Type1}, we measure Type-I correlation of the array domain channels and frequency domain channels to investigate the correlation between the sub-data in AL-AD and AL-FD, respectively, and to choose an appropriate channel domain for AL.
Using the definition of the covariance function between complex random vectors $\ba$ and $\bb$,
\begin{align}
	\text{cov}\left( \ba, \bb \right) = \mathbb{E} \left[\left(\ba- \mathbb{E} \left[\ba \right] \right)^\mathrm{H} \left( \left[ \bb- \mathbb{E} \left[ \bb \right]\right] \right)  \right], 
\end{align}
Type-I correlation between the $i$-th and $i^\prime$-th sub-channels is computed as 
\begin{align}\label{eq: eq14}
	r_\text{I}\left(i, i^\prime \right)=\frac{\text{cov}\Big(\bH_{n}^i[:], \bH_{n}^{i^\prime}[:]\Big)}{\sqrt{\text{cov}\Big(\bH_{n}^i[:], \bH_{n}^{i}[:]\Big)} \sqrt{\text{cov}\Big(\bH_{n}^{i^\prime}[:], \bH_{n}^{i^\prime}[:]\Big)}}.
\end{align}

In Fig. \ref{fig: Type1}, we investigate Type-I correlation of the array domain channels and frequency domain channels using the channel with UE mobility of 20 km/h while other parameters are the same as in Section \ref{Numerical}.
The expectation values are computed by averaging the sampled covariance values over 100 time slots. 
In Fig. \ref{fig: Type1} (a), Type-I correlation values are relatively high among the array domain channels, where the average value across all subcarrier indices ranges from 0.6 to 0.8. This level of correlation suggests that the training dataset for AL-AD may exhibit considerable redundancy. On the contrary, Type-I correlation values in Fig. \ref{fig: Type1} (b) are only high when the antenna pairs are close to each other in a physical sense, where the average value over all antenna indices ranges from 0.01 to 0.02. The periodical pattern is due to a uniform planar array (UPA) structure of the BS antennas and a uniform linear array (ULA) structure of UE antennas. Since frequency domain channels are less correlated to each other compared to array domain channels, it is expected that AL-FD can further improve the quality of the training dataset by reducing the redundancy. These results clearly show that Type-I correlation is indeed an effective tool within the AL approach to analyze the diversity of the aggregated training data.

\subsection{TYPE-II CORRELATION}

\begin{figure}[]
	\subfloat[Correlation between $M$ elements in the array domain channels. ]{{\includegraphics[width=0.5\textwidth ]{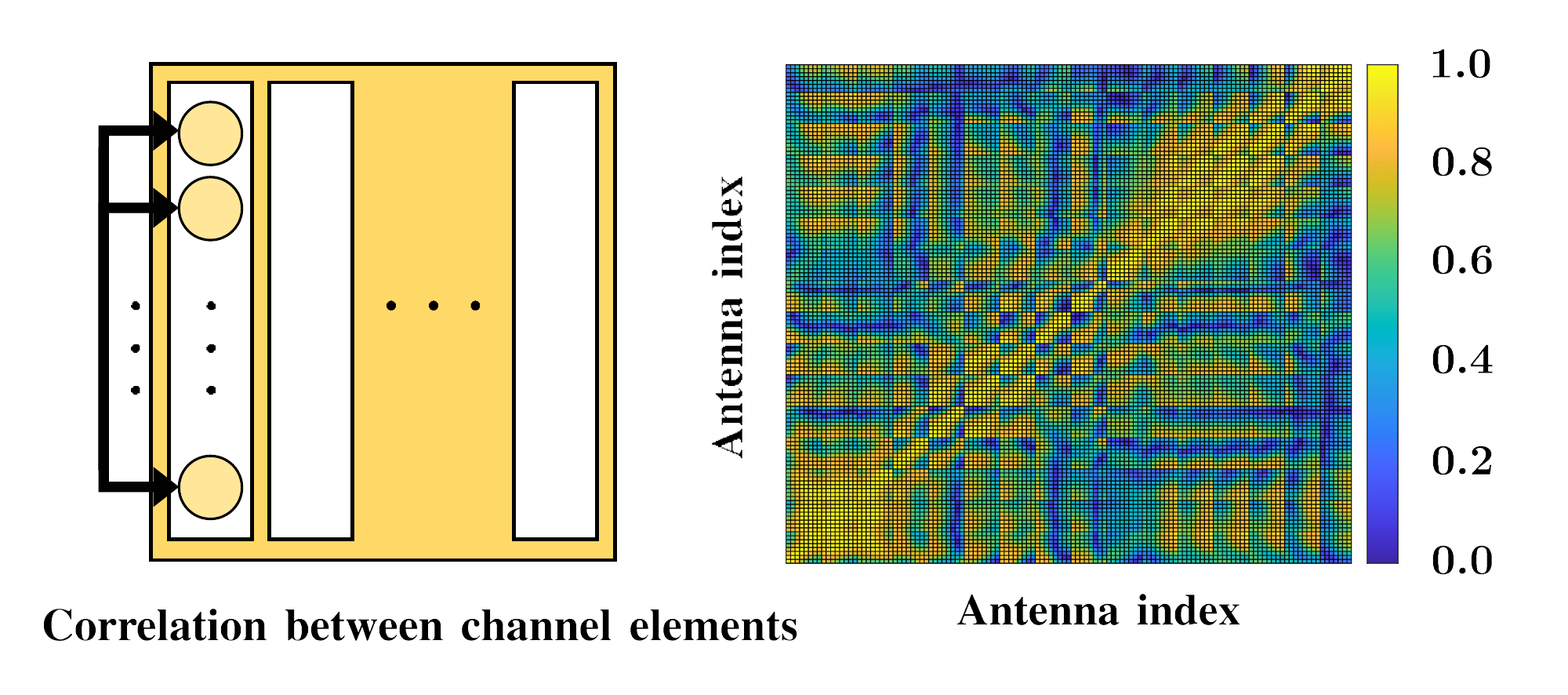} }} 		
	\centering			
	\vfill
	\subfloat[Correlation between $L$ elements in the frequency domain channels.]{{\includegraphics[width=0.5\textwidth ]{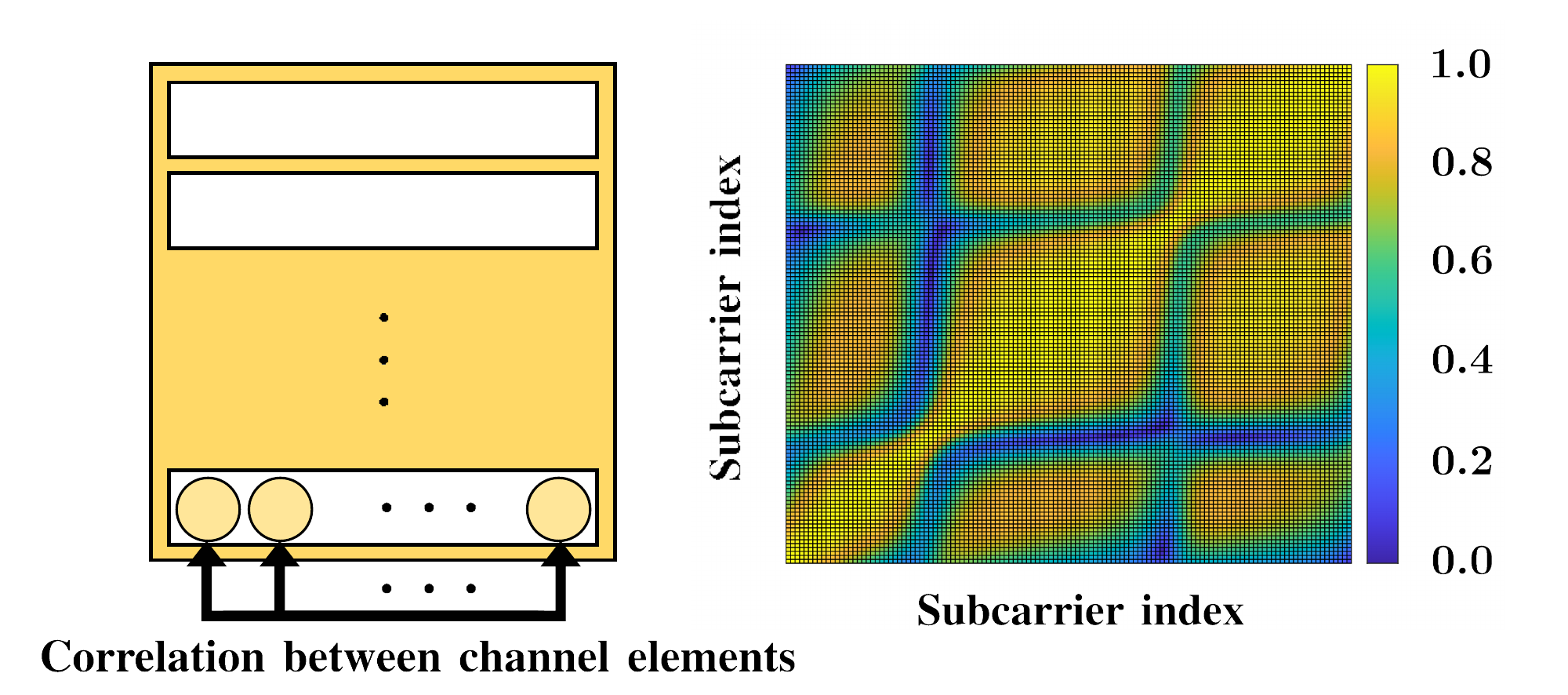} }}
	\centering		
	\caption{Type-II correlation of the array domain channels and frequency domain channels.} 		
	\label{fig: Type2}	
\end{figure}

Type-II correlation refers to the element-wise correlation within a vector channel, which can be measured by calculating the correlation between the elements of the vector, as illustrated in Fig. \ref{fig: Type2}.
While aggregating multiple sub-data from different channel domains makes the difference in the diversity of the training dataset in AL-AD and AL-FD, splitting the original training data in different channel domains makes the distinct composition of the sub-data in each training dataset of AL-AD and AL-FD.
Specifically, the sub-data in AL-AD only consists of array domain channels and the sub-data in AL-FD only consists of frequency domain channels. 
Hence, we investigate Type-II correlation of the array domain channels and frequency domain channels. Type-II correlation of the $i$-th sub-channel is computed as
\begin{align}
	r_\text{II}^i\left(j, j^\prime \right)=\frac{\text{cov}\Big(\bH_n^i[j], \bH_n^i[j^\prime]\Big)}{\sqrt{\text{cov}\Big(\bH_n^i[j], \bH_n^i[j]\Big)} \sqrt{\text{cov}\Big(\bH_n^i[j^\prime], \bH_n^i[j^\prime]\Big)}}.
\end{align}

In Fig. \ref{fig: Type2}, we examine Type-II correlation for both array and frequency domain channels, using the same channel described in Section \ref{Correlation}-\ref{Correlation-A}. We compute the average of Type-II correlation values across $L$ subcarriers for array domain channels and across $M$ antennas for frequency domain channels.
We observe that Type-II correlation values of the frequency domain channels are higher than those of the array domain channels, which implies that the frequency domain channels are more spatially correlated than the array domain channels. 
Hence, the neural networks for AL-AD and AL-FD will explore different levels of Type-II correlation, and the impact of these differences will be made clear in Section \ref{Numerical}.

\subsection{TEMPORAL CORRELATION}

Considering the dependency of performance of ML-based channel predictors on temporal correlation and recognizing that a strong temporal correlation is advantageous for channel prediction, we compute the temporal correlation of the array domain channels and frequency domain channels. The temporal correlation of the $i$-th sub-channel can be computed as
\begin{align}
	R^{i}(k)= \mathbb{E} \left[ \frac{( {\bH_n^i[:]} )^\mathrm{H} \bH_{n+k}^i[:]}{\lVert \bH_n^i[:] \rVert^2} \right],
\end{align}
where $k$ is the time shift. 

\begin{figure}[]
	\centering
	\includegraphics[width=1.0\columnwidth]{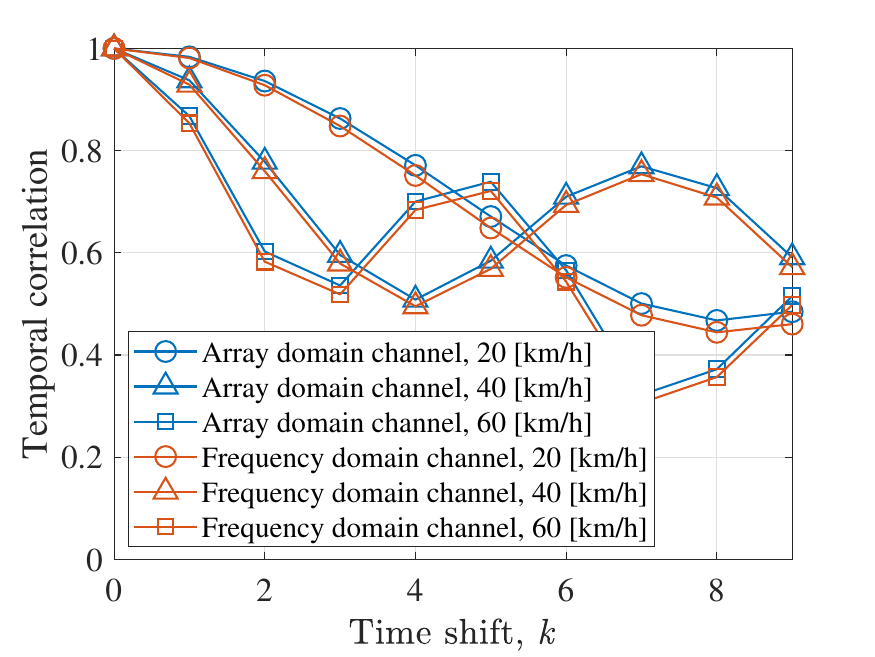}
	\caption{Temporal correlation of the array domain channels and frequency domain channels with various UE mobilities.}\label{fig: Type3}
\end{figure}

In Fig. \ref{fig: Type3}, we examine the temporal correlation of both the array domain and frequency domain channels, utilizing the channels in which UE mobility is set at 20, 40, and 60 km/h. We take the average of the temporal correlation values for the array domain channels and frequency domain channels over $L$ subcarriers and $M$ antennas, respectively.
The analysis shows that the temporal correlation undergoes more rapid fluctuations with increasing mobility of the UE across both the array and frequency domain channels. Despite the fluctuation due to the UE mobility, the difference in temporal correlation between the array and frequency domain channels is not significant compared to differences in Type-I correlation and Type-II correlation. Hence, we can expect that Type-I correlation and Type-II correlation will play critical roles for the prediction performance in AL-AD and AL-FD.

\section{BENCHMARKS}\label{Benchmarks}
To evaluate the efficacy of the data pre-processing in the AL approach, we introduce several benchmarks within the online re-training framework. First, we present a separate learning (SL) approach, which only involves splitting the original training data and separately trains neural networks to perform predictions for each sub-channel. We also consider the SL approach with data augmentation and meta-learning. Second, we introduce a joint learning (JL) approach, which predicts the channel without any data pre-processing, directly using the original MIMO-OFDM channels as training data to exploit spatio-temporal correlations (both array and frequency domains jointly).

\subsection{SL APPROACH}
\begin{figure}[]
	\centering
	\includegraphics[width=0.99\columnwidth]{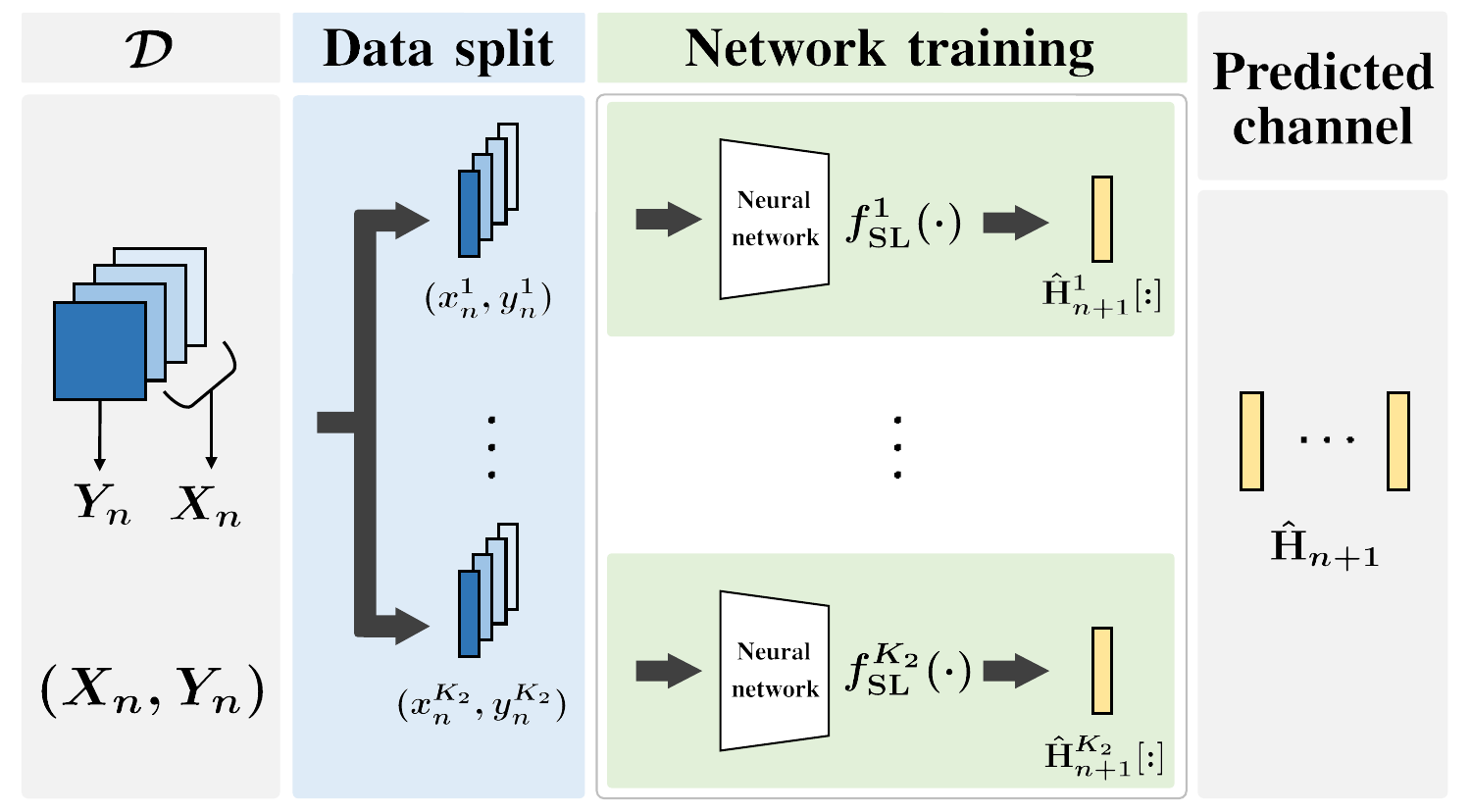}
	\caption{Overall process of the SL approach.}\label{fig: SL}
\end{figure}

The SL approach only splits the training data $\left(X_n, Y_n \right)$ into sub-data for each sub-channel, while the AL approach further processes the training data by aggregating multiple sub-data into a new training dataset. Following the data split, the SL approach performs independent channel prediction for each sub-channel as described in Fig. \ref{fig: SL}.

The training dataset for the $i$-th sub-channel in the SL approach is denoted as  
\begin{align}
	\mathcal{D}_\text{SL}^i = \left\{ \left(x_n^i, y_n^i \right) | n \in \mathcal{N}_\text{tr} \right\}, 
\end{align}
where $| \mathcal{D}_\text{SL}^i |=N-I$.
For each sub-channel, the SL approach employs the same neural network architecture and hyper-parameters, including learning rate, batch size, and number of epochs, as those used in the AL approach. This ensures that the evaluation is concentrated on the data pre-processing aspect of the AL approach. The loss function for the $i$-th neural network is given as
\begin{align}
	\text{Loss}_\text{SL}^i=\frac{1}{|\mathcal{N}_\text{tr}|}\sum_{n \in \mathcal{N}_\text{tr}} \left\Vert \bG_{n+1}^i[:] - \hat{\bH}_{n+1}^i[:] \right\Vert_2^2.
\end{align} 
The channel predictor for the $i$-th sub-channel is generated after training the $i$-th neural network with $\mathcal{D}^i_\text{SL}$, and the channel prediction for the $i$-th sub-channel is formulated as 
\begin{align}
	\hat{\bH}_{n+1}^i[:] = f_{\text{SL}}^i\left(\left\{\bG_k^i[:] \right\}_{k=n-I+1}^n \right), \ n \in \mathcal{N}_\text{pr}.
\end{align}
By repeating the training and prediction for $K_2$ sub-channels, the array-frequency domain channel for the MIMO-OFDM system can be reconstructed.
Since there are two types of sub-channels, i.e., the array domain and frequency domain channels, there are two distinct variants for the SL approach, SL in the array domain (SL-AD) and SL in the frequency domain (SL-FD). 
The SL approach, especially in the array domain, can be interpreted as employing an ML-based channel predictor for narrowband MIMO systems, e.g., as in \cite{MLP} adopting MLP for the neural network, for each sub-channel separately.

\textit{Remark 2: }Compared to the SL approach, the AL approach requires about $1/K_2$ of the number of time slots for data collection to achieve the comparable amount of training data. Moreover, the AL approach reduces the computational overhead in network training by generating a single predictor for the array-frequency domain channel prediction as clearly shown in Fig. \ref{fig: AL}, whereas the SL approach requires generating $K_2$ predictors for the same task as in Fig. \ref{fig: SL}.

To further compare the AL approach with ML techniques commonly employed under conditions of limited number of training data, we separately incorporated data augmentation and meta-learning into the SL approach. Data augmentation effectively enlarges the dataset by artificially increasing the number of training samples \cite{DATAAUG1, DATAAUG2}. In contrast, meta-learning optimizes the learning process to make efficient use of a small dataset. Specifically, it prepares neural networks to quickly adapt to new tasks using minimal data, utilizing strategies developed during the meta-training phase \cite{Timothy2022}.
\begin{itemize}
	\item SL X FLIP: To compare the AL approach with conventional techniques under the limited training dataset conditions, we integrate data augmentation into the SL approach. Specifically, we implement the flipping technique detailed in \cite{DATAAUG3}. For the $i$-th sub-channel, the training dataset is augmented with its vertically flipped version, effectively doubling the amount of training data. In the SL-AD with flipping (SL-AD X FLIP), the training dataset includes instances where the antenna components are flipped. Similarly, in the SL-FD with flipping (SL-FD X FLIP), the training dataset includes instances where the subcarrier components are flipped.
	\item SL X MAML: A channel predictor utilizing meta-learning, described in \cite{Kim2023}, is integrated into the SL approach, leveraging model-agnostic meta-learning (MAML), which is a prominent meta-learning algorithm \cite{Finn2017}. Unlike standard SL, where network parameters start from random initialization, SL X MAML begins with parameters pre-trained during a meta-training stage using additional data from other UEs as detailed in \cite{Kim2023}. This pre-training facilitates efficient adaptation for a new environment with a limited training dataset. Specifically, SL X MAML adapts this pre-trained network for the $i$-th sub-channel using its corresponding dataset $\mathcal{D}_\text{SL}^i$ in the meta-adaptation stage. Furthermore, SL-AD and SL-FD variants integrated with MAML are denoted as SL-AD X MAML and SL-FD X MAML, respectively.
\end{itemize}

\subsection{JL APPROACH}
\begin{figure}[]
	\centering
	\includegraphics[width=1.0\columnwidth]{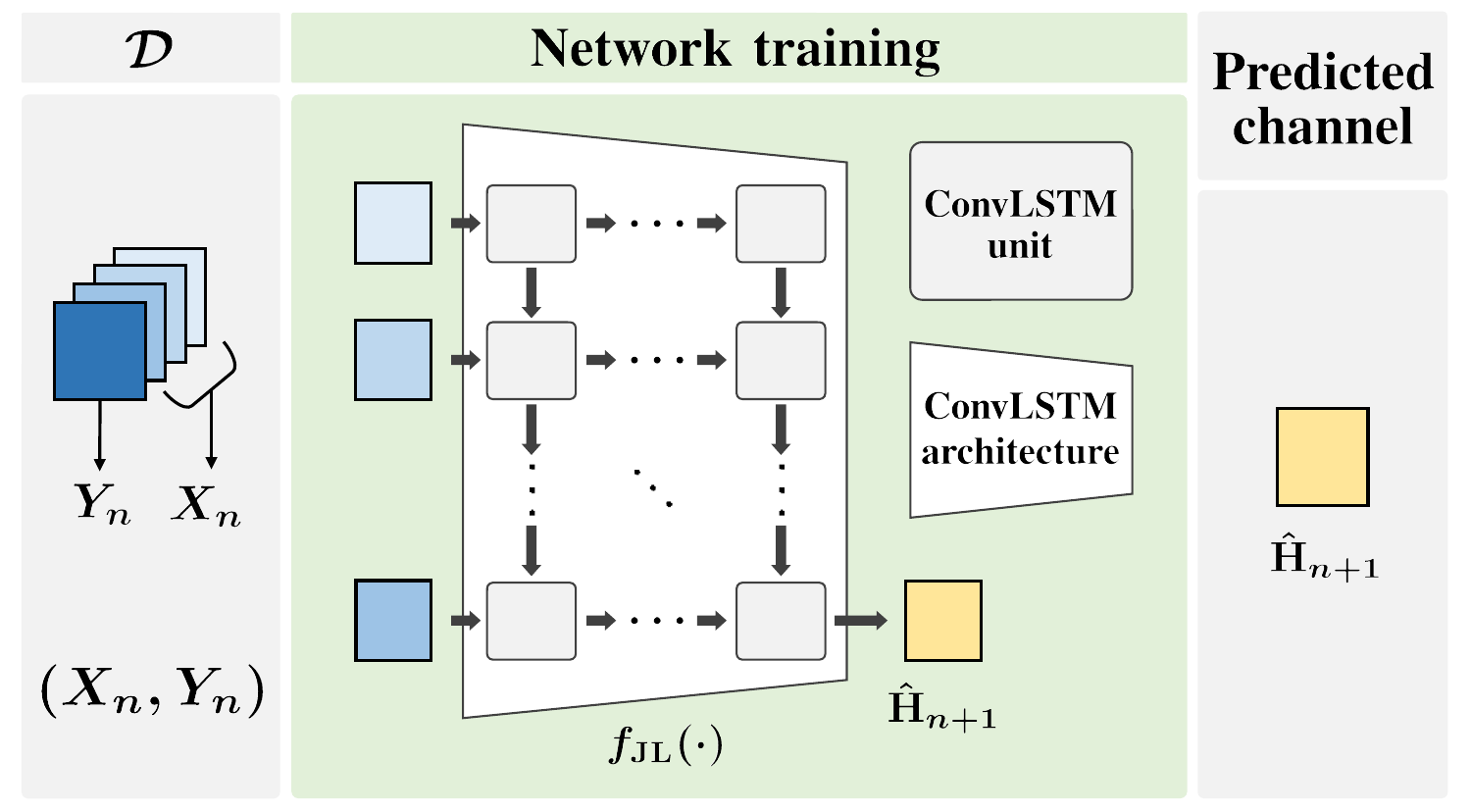}
	\caption{Overall process of the JL approach.}\label{fig: JL}
\end{figure}
We also present the JL approach, which does not pre-process the training data $\left(X_n, Y_n \right)$, but directly inputs into the neural network as in conventional ML-based channel prediction for MIMO-OFDM systems \cite{OFDM2}. As shown in Fig. \ref{fig: JL}, the JL approach uses the ConvLSTM architecture, specifically chosen for its suitability in capturing spatio-temporal correlations of the time-varying two-dimensional data \cite{ConvLSTM}. Unlike the AL and SL approaches, which split the training data into multiple sub-channels, the JL approach jointly learns the spatio-temporal correlations of both the array and frequency domains, as noted in \textit{Remark 1}. The loss function of the JL approach is given as 
\begin{align}
	\text{Loss}_\text{JL}=\frac{1}{|\mathcal{N}_\text{tr}|}\sum_{n \in \mathcal{N}_\text{tr}} \left\Vert \bG_{n+1} - \hat{\bH}_{n+1} \right\Vert_2^2,
\end{align} 
and the channel prediction for the JL approach is formulated as 
\begin{align}
	\hat{\bH}_{n+1}=f_\text{JL}\left(\left\{\bG_k \right\}_{k=n-I+1}^n \right), \ n \in \mathcal{N}_\text{pr},
\end{align}
where the predicted array-frequency domain channel is directly provided from the output.

\section{NUMERICAL RESULTS}\label{Numerical}

\subsection{SIMULATION PARAMETERS}
This paper utilizes the quasi-deterministic radio channel generator (QuaDRiGa) in \cite{QUA} to generate channels, with consideration of an urban micro (UMi) scenario. The simulation parameters include a carrier frequency of 2.53 GHz, the time slot duration of 2 ms, and the configuration of 128 subcarriers with each subcarrier spacing of 40 kHz. At the BS, an $8 \times 8$ UPA with an antenna spacing of half a wavelength is implemented, consisting of a total of 64 antennas. The UE is equipped with a ULA with 2 antennas, also spaced half a wavelength apart.  
For the channel estimation, the pilot power is set to $\rho=10$ dBm, and the pilot length $\tau$ is fixed at 2 to achieve column-wise orthogonality in the pilot matrix, utilizing a discrete Fourier transform (DFT) matrix. Additionally, the noise variance $\sigma^2$ is determined based on the noise spectral density of -174 dBm/Hz.

For the network architectures in the AL and SL approaches, we primarily employ an MLP for demonstration purposes unless stated otherwise, while comparisons with RNN, LSTM, and transformer are also conducted.
The MLP is equipped with two dense layers, each containing $N_\text{node}=2IK_1$ nodes. A rectified linear unit (ReLU) serves as the activation function. 
For the JL approach using ConvLSTM, a single layer is implemented with a filter size of one and a kernel size of $16 \times 16$. The recurrent activation function is set to a sigmoid function, and the activation function is set to a hyperbolic tangent function. Key hyper-parameters include a batch size of 16, a learning rate of 0.001, and a training duration of $N_\text{epoch}$ = 150 epochs.

We evaluate the performance of channel prediction using the normalized MSE (NMSE), which is defined as
\begin{align}
	\text{NMSE}=\mathbb{E} \left[ \frac{\lVert{ \bH_{n+1} - {\hat{\bH}}_{n+1} }\rVert^2_\text{F}} {  \lVert{{{\bH}}_{n+1} }\rVert^2_\text{F} } \right].
\end{align}	
The prediction performance is tested for $100$ time slots, and the time gap between the set of time slots for training $\mathcal{N}_\text{tr}$ and prediction $\mathcal{N}_\text{pr}$ is fixed at 100 time slots. Also, we evaluate the achievable sum-rate assuming $U$ single antenna UEs in the system. The online re-training framework operates in cycles of duration $T_\text{cyc}$, and during each cycle, outdated estimated channels and predicted channels are used for data transmission in the training phase and prediction phase, respectively.
To mitigate the inter-user interference, a zero-forcing (ZF) combiner is employed as in \cite{Kim2023, Tao2021}. During the training phase, the ZF combiner for the $\ell$-th subcarrier is given as
\begin{align}
	{\bF_{n, \text{tr}}^{\ell}}^\mathrm{T}=\left({{}{\bar{\bG}}_n^\ell}^\mathrm{H} {{}{\bar{\bG}}_n^\ell}\right)^{-1} {{}{\bar{\bG}}_n^\ell}^\mathrm{H},
\end{align}
where ${{\bar{\bG}}_n^\ell}=[\bG_n^1[:, \ell], \cdots, \bG_n^U[:, \ell]] \in \mathbb{C}^{M \times U}$ represents the channel state information for the $\ell$-th subcarrier, with $\bG_n^u$ denoting the estimated array-frequency domain channel of the $u$-th UE.
Similarly, during the prediction phase, the ZF combiner for the $\ell$-th subcarrier is given as
\begin{align}
	{\bF_{n, \text{pr}}^\ell}^\mathrm{T}=\left({{}\bar{\bH}_n^\ell}^\mathrm{H} {{}\bar{\bH}_n^\ell}\right)^{-1} {{}\bar{\bH}_n^\ell}^\mathrm{H},
\end{align}
where ${\bar{\bH}_n^\ell}=[\hat{\bH}_{n+1}^1[:, \ell], \cdots, \hat{\bH}_{n+1}^U[:, \ell]] \in \mathbb{C}^{M \times U}$ represents the predicted channel state information, with $\hat{\bH}_{n+1}^u$ denoting the predicted array-frequency domain channel of the $u$-th UE. For each UE, the combiner is normalized as ${\bar{\bff}_{n, \text{tr}}^{\ell, u}}={{\bff}_{n, \text{tr}}^{\ell, u}}/{\lVert {\bff}_{n, \text{tr}}^{\ell, u} \rVert}$ for the training phase and $\bar{\bff}_{n, \text{pr}}^{\ell, u}={{\bff}_{n, \text{pr}}^{\ell, u}}/{\lVert {\bff}_{n, \text{pr}}^{\ell, u} \rVert}$ for the prediction phase. Here, ${\bff}_{n, \text{tr}}^{\ell, u}$ and ${\bff}_{n, \text{pr}}^{\ell, u}$ represent the $u$-th column of ${\bF_{n, \text{tr}}^\ell}$ and ${\bF_{n, \text{pr}}^\ell}$, respectively.
The achievable rate of the $u$-th UE in the training phase is given as
\begin{align}
	R^u_\text{tr}=\frac{1}{L}\sum_{\ell=1}^L\alpha \log \left(1+ \frac{\gamma \lvert {{}\bar{\bff}_{n, \text{tr}}^{\ell, u}}^\mathrm{T} {{\bH}_{n+1}^u[:, \ell]} \rvert^2}{\gamma \sum_{v\ne u}\lvert {{}\bar{\bff}_{n, \text{tr}}^{\ell, u}}^\mathrm{T}  {{\bH}_{n+1}^v[:, \ell]} \rvert^2+\sigma^2} \right),
\end{align}
and in the prediction phase, the rate is given as
\begin{align}
	R^u_\text{pr}=\frac{1}{L}\sum_{\ell=1}^L\alpha \log \left(1+ \frac{\gamma \lvert {{}\bar{\bff}_{n, \text{pr}}^{\ell, u}}^\mathrm{T} {{\bH}_{n+1}^u[:, \ell]} \rvert^2}{\gamma \sum_{v\ne u}\lvert {{}\bar{\bff}_{n, \text{pr}}^{\ell, u}}^\mathrm{T}  {{\bH}_{n+1}^v[:, \ell]} \rvert^2 + \sigma^2} \right),
\end{align}
where ${\bH}_{n+1}^u$ is the true array-frequency domain channel of the $u$-th UE, $\gamma$ is the transmit signal power, and $\alpha=(N_s-\tau)/N_s$.
The overall achievable sum-rate is given as
\begin{align}
	R_\text{sum}=\sum_{u=1}^U \beta R^u_\text{tr} + (1-\beta) R^u_\text{pr},
\end{align}
where $\beta={T_\text{tot}}/{T_\text{cyc}}$.

\subsection{PREDICTION PERFORMANCES}
\begin{figure}[t]
	\centering
	\includegraphics[width=1.05\columnwidth]{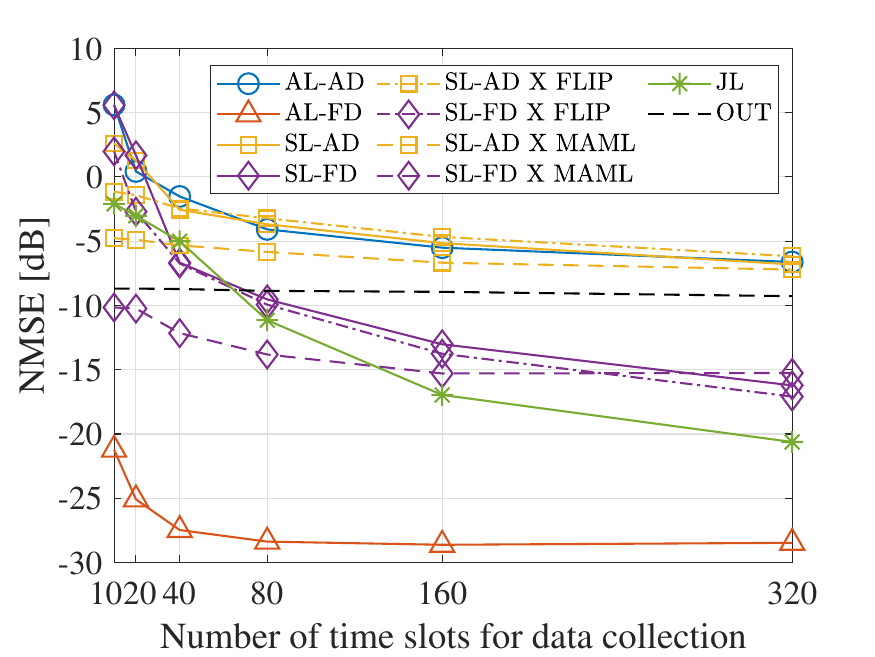}
	\caption{NMSE performances of AL-AD, AL-FD, SL-AD, SL-FD, SL-AD and SL-FD with FLIP and MAML, and JL with respect to number of time slots for data collection $N$.}\label{fig: NMSE_data}
\end{figure}

The prediction performances of AL-AD, AL-FD, SL-AD, SL-FD, SL-AD and SL-FD with FLIP and MAML, as well as JL, are examined across varying number of time slots for data collection $N$ in Fig. \ref{fig: NMSE_data}. For SL X MAML, we utilize the same network architecture as that employed in the AL and SL approaches to ensure consistency. The meta-training stage employs training data from four different UEs, collected across 160 time slots, and the inner-task learning rate is set at 0.1, while the outer-task learning rate remains at 0.001. The UE mobility is set to 20 km/h, and the input order $I$ is fixed at two. 
The graph labeled OUT represents the NMSE values evaluated with the outdated estimated channels, showing the quality of CSI without any channel prediction.
As depicted in the figure, the prediction performances of all predictors improve as the number of time slots for data collection is doubled. This implies the importance of training ML-based channel predictors with sufficient amount of training data to ensure reliable performance. Note that the NMSE value of OUT remains stable, as channel estimation is independent of channel prediction.
Among all approaches, AL-FD demonstrates the best performance, while AL-AD exhibits a notable performance gap compared to AL-FD and shows similar performance to SL-AD, which does not perform data aggregation. When comparing the AL and SL variants across channel domains, i.e., array and frequency domains, frequency domain variants generally outperform those in the array domain. For SL X FLIP, SL-AD X FLIP performs slightly worse than SL-AD, and SL-FD X FLIP shows a slight improvement over SL-FD. The performance variations are minimal since FLIP simply adds duplicates of the training data by flipping. However, for SL X MAML, both SL-AD X MAML and SL-FD X MAML exhibit performance gains, especially when the number of time slots is limited. In the JL approach, while it achieves similar performance to SL-FD with a limited number of training data, its performance exceeds all other approaches except AL-FD when training data is more abundant.

Considering the motivation of the AL approach is to ensure prediction performance despite constraints on data collection time, AL-FD demonstrates superior performance compared to other predictors, even under the restriction of only 10 time slots for data collection. The superior efficacy of AL-FD can be attributed to the synergy of the AL approach's data pre-processing and distinctive correlation properties exhibited by frequency domain channels in MIMO-OFDM systems. These correlation properties include low Type-I correlation and high Type-II correlation. 

\begin{table}[]
	\renewcommand{\arraystretch}{1.5}
	\centering
	\caption{Correlation levels of the array domain channels and frequency domain channels with different antenna spacing.}
	\label{tab: Table1}
	\begin{tabular}{cccc}
		\hline \hline
		\multirow{2}{*}{Antenna spacing} & \multirow{2}{*}{Channel domain} & \multicolumn{2}{c}{Correlation type} \\ \cline{3-4} 
		&  & Type-I & Type-II \\ \hline \hline
		\multirow{2}{*}{0.1$\lambda$} & Array domain & 0.67 & 0.71 \\ \cline{2-4} 
		& Frequency domain & 0.71 & 0.68 \\ \hline
		\multirow{2}{*}{0.5$\lambda$} & Array domain & 0.64 & 0.01 \\ \cline{2-4} 
		& Frequency domain & 0.01 & 0.66 \\ \hline
		\multicolumn{2}{c}{Preferred correlation level for AL} & Low & High \\ \hline \hline
	\end{tabular}
\end{table}

Initially, the influence of Type-I correlation on the AL approach is evident when comparing the predictive performance of AL-AD and AL-FD. With data aggregation, the total amount of training data increases by $L$-fold for AL-AD and $M$-fold for AL-FD compared to SL-AD and SL-FD, respectively. However, only AL-FD exhibits performance improvement from increased number of training data, while AL-AD does not show any enhancement compared to SL-AD. This discrepancy is related to Type-I correlation among sub-channels, influencing redundancy in the dataset, as detailed in Section \ref{Correlation}. Specifically, the expanded dataset in AL-AD does not yield the same benefits as in AL-FD due to the high correlation among array domain channels, leading to excessive redundancy in the training dataset compared to AL-FD. This suggests that the efficacy of the AL approach is enhanced when the sub-channels are less correlated.

Furthermore, by comparing SL-AD and SL-FD, we analyze the impact of Type-II correlation on prediction performance, noting that both SL-AD and SL-FD are unaffected by Type-I correlation and share similar temporal correlation. As mentioned in Section \ref{Correlation}, the array domain channels exhibit low Type-II correlation, whereas frequency domain channels experience high Type-II correlation. Hence, the observation that SL-FD has better performance than SL-AD suggests that higher Type-II correlation significantly enhance the performance of ML-based channel predictions. Consequently, the superior performance of AL-FD can be also attributed to its frequency domain channels in the training dataset, which are marked by Type-II correlation.

To further validate the argument that the performance enhancement of AL-FD is related to low Type-I correlation and high Type-II correlation of the frequency domain channels, we analyze the prediction performance with varying antenna spacing. Such spacing adjustments affect Type-I correlation and Type-II correlation, enabling a deeper understanding of the impact on the performance of the AL approach. As described in Table \ref{tab: Table1}, for the array domain channels, Type-I correlation is relatively stable despite varying spacing since the correlation among array domain channels from different subcarriers is not affected by antenna spacing. On the contrary, Type-II correlation decreases as spacing increases since the correlation among the elements in the array domain channels actually decreases from increased spacing. Meanwhile, for frequency domain channels, Type-I correlation reduces with increased spacing, reflecting reduced correlation among frequency domain channels from different antennas, whereas Type-II correlation remains consistent due to the fact that each element of frequency domain channels are only represented with channel values from different subcarriers. 

\begin{figure}[t]
	\centering
	\includegraphics[width=1.05\columnwidth]{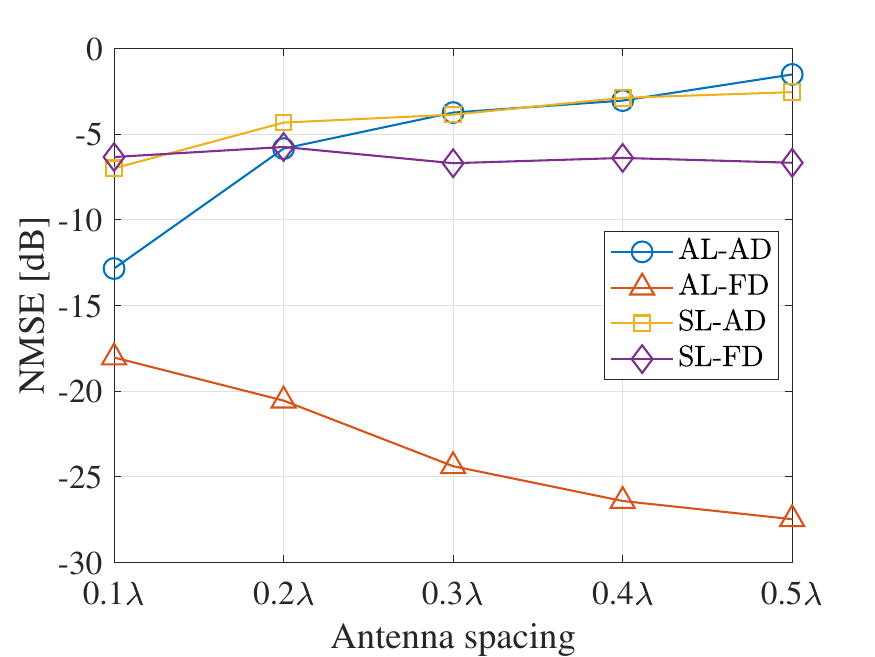}
	\caption{NMSE performance of AL-AD, AL-FD, SL-AD, and SL-FD with respect to antenna spacing.}\label{fig: NMSE_ant}
\end{figure}

Given the correlation levels presented in Table \ref{tab: Table1}, we investigate the prediction performances of AL-AD, AL-FD, SL-AD, and SL-FD with respect to varying antenna spacing in Fig. \ref{fig: NMSE_ant}. The analysis is based on the channel with UE mobility of 20 km/h, adjusting the antenna spacing from $0.1 \lambda$ to $0.5 \lambda$ with $0.1 \lambda$ increments, where $\lambda$ represents the wavelength.
The input order is set to two, and the number of time slots for data collection $N$ is fixed at 40. Initially, we examine the performance changes in the SL approach, which are unaffected by Type-I correlation but influenced by Type-II correlation. For SL-AD, an increase in antenna spacing leads to performance degradation, which is attributed to the effect of decreasing Type-II correlation. For SL-FD, the performance remains stable, which results from the constancy of Type-II correlation despite increase in antenna spacing. Next, we delve into the AL approach, which is impacted by both Type-I correlation and Type-II correlation. In the case of AL-AD, performance deteriorates with an increase in antenna spacing, similar to the trend observed in SL-AD, which is due to the decrease in Type-II correlation. However, for AL-FD, prediction accuracy improves as antenna spacing increases. Since Type-II correlation does not significantly vary for AL-FD, it can be deduced that low Type-I correlation positively affects the prediction performance of the AL approach. In summary, as observed from Figs. \ref{fig: NMSE_data} and \ref{fig: NMSE_ant}, low Type-I correlation leads to non-redundant training data, thereby enhancing the prediction performance of the AL approach. Note that while the AL approach may appear to sacrifice partial correlation of the other domain due to splitting, it still exploits this correlation as Type-I correlation, which contributes to the efficiency of the aggregated sub-data in the new training dataset. Moreover, high Type-II correlation provides advantage for both AL and SL approaches.

\begin{figure}[t]
	\centering
	\includegraphics[width=1.05\columnwidth]{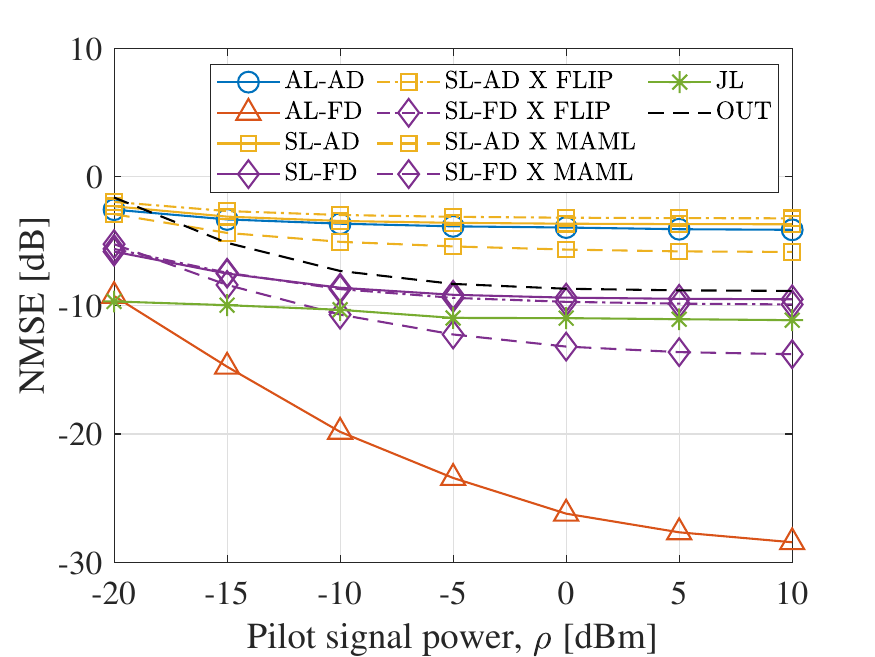}
	\caption{NMSE performances of AL-AD, AL-FD, SL-AD, SL-FD, SL-AD and SL-FD with FLIP and MAML, and JL with respect to pilot signal power.}\label{fig: NMSE_power}
\end{figure}

The prediction performances of AL-AD, AL-FD, SL-AD, SL-FD, and their combinations with FLIP and MAML, as well as JL, are analyzed with respect to the pilot signal power in Fig. \ref{fig: NMSE_power}. The UE mobility is set to 20 km/h, the input order $I$ is fixed at two, and the number of time slots for data collection $N$ is set to 80. As the pilot signal power increases, the prediction performance improves across every predictor. While SL-FD shows slightly better performance than OUT, both AL-AD and SL-AD exhibit the worst performance, even worse than OUT, due to a scarcity of adequate training data and low Type-II correlation of the array domain channels. 
Similar to the standard SL-AD, SL-AD X FLIP and SL-AD X MAML show performances that are worse than OUT, even with the use of data augmentation and meta-learning, respectively. Conversely, SL-FD X FLIP and SL-FD X MAML demonstrate performance gains, with SL-FD X MAML showing a larger gap as the pilot signal power increases. Meanwhile, the JL approach exhibits a slow decay in NMSE performance relative to the pilot signal power.
Notably, there is a significant gap between AL-FD and the other predictors, and this gap widens as the pilot signal power increases, indicating the superior effectiveness of AL-FD in exploiting the pilot signal power compared to other predictors.

\begin{figure}[t]
	\centering
	\includegraphics[width=1.05\columnwidth]{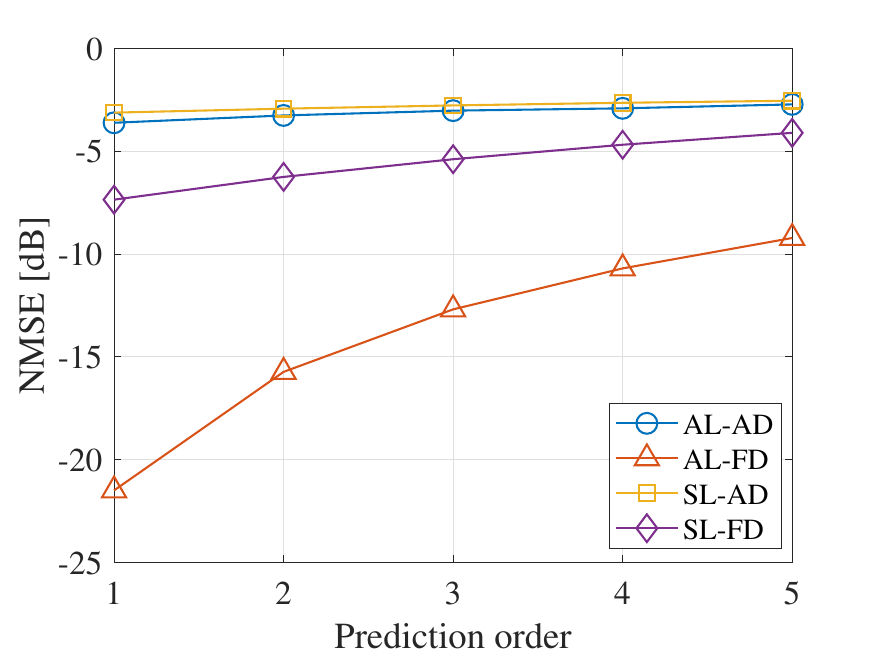}
	\caption{NMSE performance of AL-AD, AL-FD, SL-AD, and SL-FD with respect to prediction order.}\label{fig: NMSE_step}
\end{figure}

While our initial focus was on the one-step ahead prediction, which predicts the channel for the immediate next time slot, addressing channel prediction for multiple subsequent steps is also crucial for channel predictors. Therefore, we investigate the accuracy of predictions over various prediction order $p$, which involve minor modifications to the structure of the training data, as explained in the footnote 3 of Section \ref{Proposed}. The UE mobility is set to 40 km/h, the input order is fixed at three, and the number of time slots for data collection $N$ is set to 40. In Fig. \ref{fig: NMSE_step}, the performances across all predictors degrade as the prediction order $p$ increases. This degradation is due to the growing temporal separation between the features and the labels. Remarkably, AL-FD consistently shows the best performance, highlighting the unique benefits of frequency domain channels for prediction, even in scenarios of limited data collection time.

\begin{figure}[t]
	\centering
	\includegraphics[width=1.05\columnwidth]{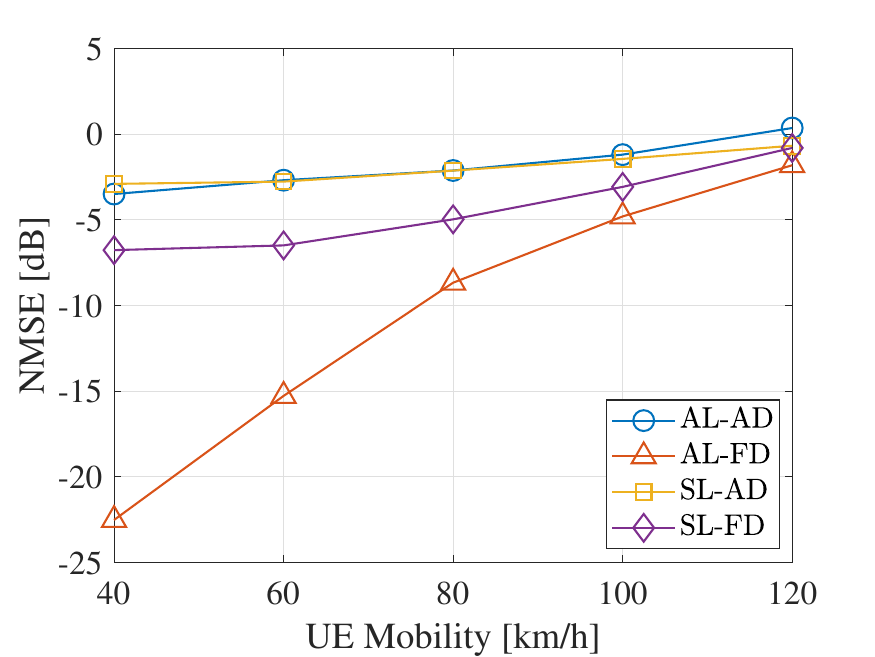}
	\caption{NMSE performance of AL-AD, AL-FD, SL-AD, and SL-FD with respect to UE mobility.}\label{fig: NMSE_speed}
\end{figure}

In Fig. \ref{fig: NMSE_speed}, we compare the performances of AL-AD, AL-FD, SL-AD, and SL-FD while increasing the mobility of the UE from 40 km/h to 120 km/h. In all cases, we kept the input order $I$ fixed at four, and the number of time slots for data collection $N$ is set to 40. Similar to varying prediction order, we observe that the performances of predictors decrease as the mobility of the UE increases. This is because the temporal correlations of both the array domain channels and frequency domain channels change more rapidly when the UE is moving fast as shown in Fig. \ref{fig: Type3}. Therefore, it is crucial to set the input order $I$ appropriately based on the mobility of UEs \cite{MLP}, as this can significantly impact the prediction performance. It is also worth noting that regardless of the UE's mobility, AL-FD outperforms AL-AD, SL-AD and SL-FD. This suggests that AL-FD may be more effective to handle the challenges posed by mobility in wireless communication systems.

\begin{figure}[t]
	\centering
	\includegraphics[width=1.05\columnwidth]{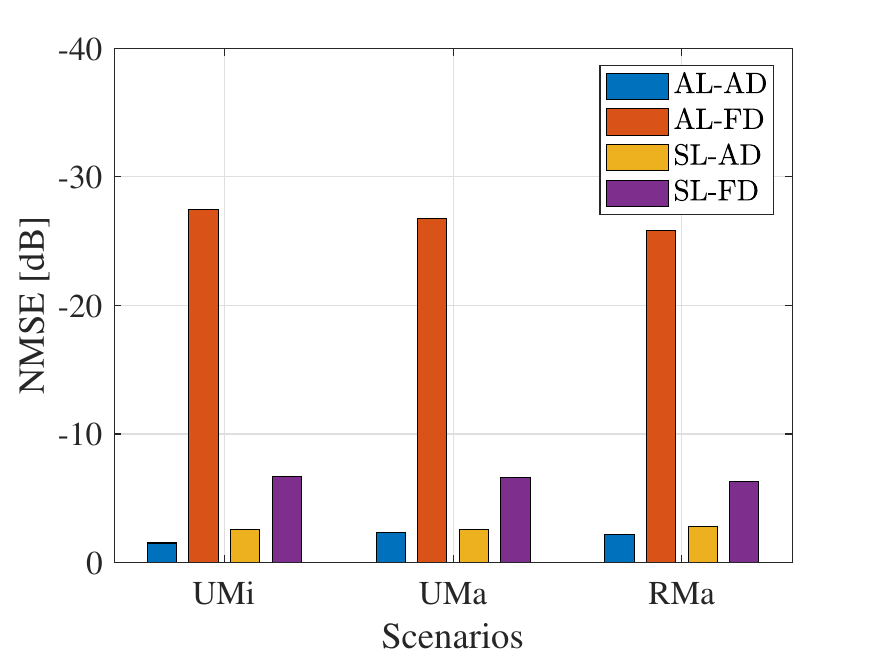}
	\caption{NMSE performance of AL-AD, AL-FD, SL-AD, and SL-FD with various channel environments.}\label{fig: NMSE_env}
\end{figure}

To check whether the proposed AL approach works for other scenarios, the NMSE performance is evaluated in three different environments, UMi, urban macro (UMa), and rural macro (RMa), as shown in Fig. \ref{fig: NMSE_env}. To ensure fair comparisons, the UE mobility is fixed at a speed of 20 km/h, the input order is set to two, and the number of time slots for data collection $N$ is set to 40 across all the environments. Note that the NMSE values are displayed in opposite orders. Although the performance gap between AL-AD, AL-FD, SL-AD, and SL-FD varies with respect to the environments, AL-FD consistently demonstrates superior prediction performance, compared to AL-AD, SL-AD, and SL-FD. The results indicate that AL-FD can be implemented not only for a specific environment but for various channel environments in the MIMO-OFDM system, and the use of the frequency domain can lead to further improvements in the prediction performance of the system in these environments.

\begin{figure}[t]
	\centering
	\includegraphics[width=1.05\columnwidth]{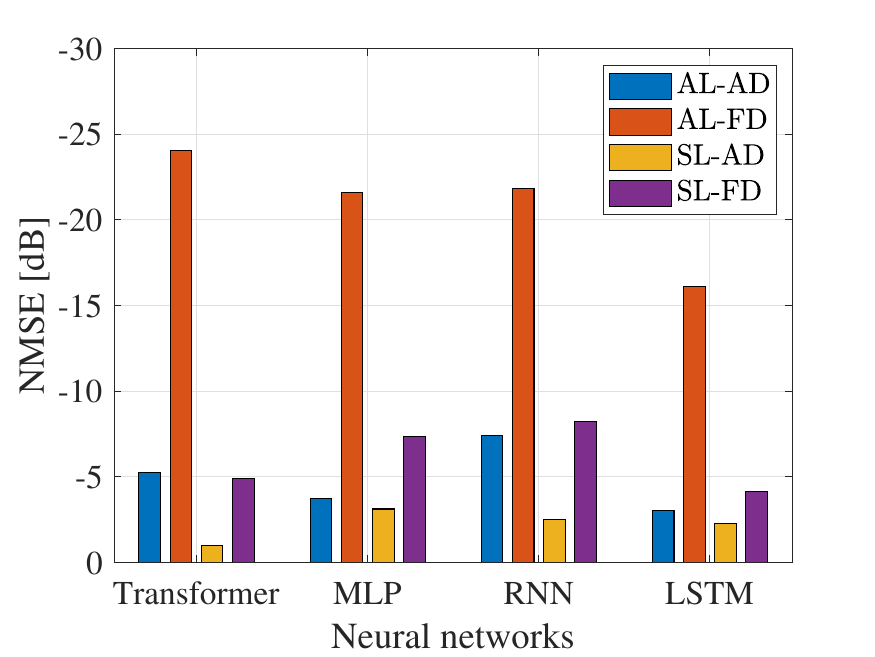}
	\caption{NMSE performance of AL-AD, AL-FD, SL-AD, and SL-FD with various neural networks.}\label{fig: NMSE_nn}
\end{figure}

In Fig. \ref{fig: NMSE_nn}, we investigate the NMSE performances of AL-AD, AL-FD, SL-AD, and SL-FD using four different neural networks: transformer, MLP, RNN, and LSTM. For the RNN and LSTM, a single layer is adopted, with each layer comprising $2K_1$ units. The sequence length is set to match the input order $I$, and a hyperbolic tangent activation function is utilized. Following this, a dense output layer with $2K_1$ nodes and linear activation is used to compile the predictions. For the transformer architecture, two encoder layers and two decoder layers are employed, with each encoder and decoder layer containing a dense layer with $2K_1$ nodes. Additionally, a multi-head attention mechanism with four heads is incorporated to capture the dependencies between the input sequences. The empirical complexities of neural networks are compared based on the trainable parameters per network in Table \ref{tab: Table2}.\footnote{While the prediction performances of transformer, MLP, RNN, and LSTM can vary depending on the choice of hyper-parameters, our primary focus in this paper is on presenting the AL approach, and determining the optimal neural network structure is beyond the scope of our current work.} The UE mobility is set to 40 km/h, input order to three, and the number of time slots for data collection to 40. Consistent with previous results, AL-FD shows the highest performance for all neural networks. Therefore, we conclude that the AL approach can be utilized with various types of neural networks, and that prediction in the frequency domain can provide extra performance gain in the MIMO-OFDM system.

\begin{table}[]
	\renewcommand{\arraystretch}{1.5}
	\renewcommand{\tabcolsep}{0.25cm}
	\centering
	\caption{Empirical evaluation of complexity for each neural network.}
	\label{tab: Table2}
	\begin{tabular}{cc}
		\hline \hline
		Neural networks & Trainable parameters per network \\ \hline \hline
		Transformer & 2,243,072  \\ \hline
		MLP & 1,378,044 \\ \hline
		RNN & 197,120 \\ \hline
		LSTM &  591,104 \\ \hline \hline 
	\end{tabular}
\end{table}

\begin{figure}[t]
	\centering
	\includegraphics[width=1.05\columnwidth]{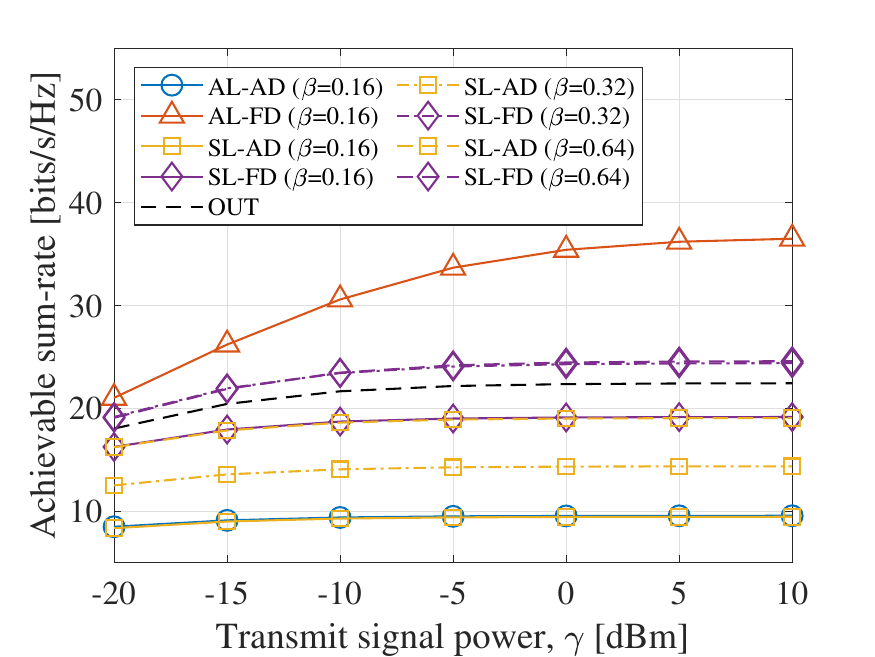}
	\caption{Achievable sum-rates of AL-AD, AL-FD, SL-AD, and SL-FD with respect to transmit signal power.}\label{fig: Sumrate}
\end{figure}

In Fig. \ref{fig: Sumrate}, we compare the achievable sum-rates of AL-AD, AL-FD, SL-AD, and SL-FD with respect to the transmit signal power $\gamma$. We set the number of UEs $U$ to 5, the number of BS antennas to 64, and the number of subcarriers $L$ to 64. The UE mobility is fixed at 20 km/h and the input order is set to two. To evaluate the effect of the proportion of training phase duration to the total duration of a cycle within the online re-training framework, we investigate three cases. First, we examine both AL and SL approaches with $\beta = 0.16$, where the number of time slots for data collection $N$ is 80. We then analyze the SL approach with $\beta = 0.32$ and $\beta = 0.64$, where $N = 160$ and $N = 320$, respectively. These cases represent increased data collection time for potentially better prediction performance, as discussed in Fig. \ref{fig: NMSE_data}. The number of symbols within each time slot $N_s$ is set to 14, resulting in $\alpha = 12/14$. For all cases, channel estimation is performed with pilot signal power $\rho = 10$ dBm.
When $\beta = 0.16$, AL-FD demonstrates the best rate performance due to superior prediction capability, while other approaches show worse rate performance than OUT. This indicates that AL-AD, SL-AD, and SL-FD exhibit poor prediction performance with $\beta = 0.16$. For $\beta = 0.32$, SL-FD achieves better performance than OUT, but SL-AD still performs worse than OUT. At $\beta = 0.64$, although SL-FD is expected to achieve much better prediction performance than at $\beta = 0.32$, the rate performance is almost similar to the $\beta = 0.32$ case. This is attributed to the reduced proportion of time utilizing the predicted channel, resulting from increased data collection time. SL-AD shows improved performance at $\beta = 0.64$ but still remains worse than OUT. We observe that to optimize rate performance, the data collection time should be minimized while maintaining sufficient prediction performance, as demonstrated by AL-FD.

\subsection{TRAINING TIME OVERHEAD AND COMPUTATIONAL COMPLEXITY}

\begin{table}[]
	\renewcommand{\arraystretch}{1.5}
	\renewcommand{\tabcolsep}{0.25cm}
	\centering
	\caption{Training time overhead, total number of training data $N_\text{train}$, and NMSE performance of AL-AD, AL-FD, SL-AD, and SL-FD.}
	\label{tab: Table3}
	\begin{tabular}{ccccc}
		\hline \hline
		\multirow{2}{*}{Predictor} & \multicolumn{2}{c}{\begin{tabular}[c]{@{}c@{}}Training time\\ overhead {[}s{]}\end{tabular}} & \multirow{2}{*}{\begin{tabular}[c]{@{}c@{}} $N_\text{train}$ \end{tabular}} & \multirow{2}{*}{NMSE {[}dB{]}} \\ \cline{2-3}
		& $T_\text{col}$ & $T_\text{com}$ &  &  \\ \hline \hline
		AL-AD& 0.02  & 14.3  &  1024 &  5.61 \\ \hline
		AL-FD& 0.02 & 14.5  &  1024  &  -21.2 \\ \hline
		\multirow{2}{*}{SL-AD} & 0.02 & 85.7 &  8 & 2.61  \\ \cline{2-5} 
		& 1.28 & 965.1 &   638 & -8.81 \\ \hline
		\multirow{2}{*}{SL-FD} & 0.02 & 86.2 &  8 & 5.58 \\ \cline{2-5} 
		& 1.28 & 954.9 &   638 & -18.5 \\ \hline \hline
	\end{tabular}
\end{table}

First, we evaluate the training time overhead, the total number of training data $N_\text{train}$, and NMSE performances of AL-AD, AL-FD, SL-AD, and SL-FD with a data collection time $T_{\text{col}} = 0.02$ seconds. Additionally, we examine for SL-AD and SL-FD with $T_{\text{col}} = 1.28$ seconds. All predictors are based on the MLP architecture. The mobility of UE is fixed at 20 km/h, and the input order is set at two. The computation time for the network training stage $T_{\text{com}}$ is measured using the graphic processing unit (GPU) capabilities of NVIDIA GeForce RTX 2080. As shown in Table \ref{tab: Table3}, AL-AD and AL-FD are capable of collecting over a thousand training data points within just 0.02 seconds, while SL-AD and SL-FD manage to collect fewer than ten in the same interval. This substantial difference underlines that SL-AD and SL-FD require more time to accumulate a comparable volume of dataset. Furthermore, the extended computation time of network training for SL-AD and SL-FD arise from the requirement to train multiple neural networks ($L$ and $M$, respectively), in contrast to the single network training required by the AL approach. Although SL-FD, with a $T_{\text{col}} = 1.28$ seconds, achieves prediction performance comparable to that of AL-FD, the latter demonstrates significantly lower training time overhead. This underscores AL-FD's efficiency in reducing training time overhead while still enhancing prediction accuracy.

\begin{table}[]
	\renewcommand{\arraystretch}{1.5}
	\renewcommand{\tabcolsep}{0.25cm}
	\centering
	\caption{Computational complexity of AL-AD, AL-FD, SL-AD, and SL-FD.}
	\label{tab: Table4}
	\begin{tabular}{cc}
		\hline \hline
		Predictor & Computational complexity \\ \hline \hline
		AL-AD & $\mathcal{O}\left(N_\text{epoch} N_\text{node} L (N-I) \left( N_\text{node} +M \left(I+1 \right) \right) \right)$  \\ \hline
		AL-FD & $\mathcal{O}\left(N_\text{epoch} N_\text{node} M (N-I)  \left( N_\text{node} +L \left(I+1 \right) \right) \right)$ \\ \hline
		SL-AD & $\mathcal{O}\left(N_\text{epoch} N_\text{node} (N-I)  \left( N_\text{node} +M \left(I+1 \right) \right) \right)$ \\ \hline
		SL-FD &  $\mathcal{O}\left(N_\text{epoch} N_\text{node} (N-I)  \left( N_\text{node} +L \left(I+1 \right) \right) \right)$ \\ \hline \hline 
	\end{tabular}
\end{table}

Second, we also analyze the computational complexity to verify the computation time of the network training. As studied in \cite{MLP}, the computational complexity of an MLP is given as
\begin{align}
	\mathcal{O}\left(N_\text{epoch} N_\text{node} N_\text{train} \left( N_\text{node} +K_1 \left(I+1 \right) \right) \right).
\end{align}
Table \ref{tab: Table4} summarizes the computational complexities of training the neural network for AL-AD and AL-FD, as well as for SL-AD and SL-FD for a single subcarrier or antenna. While the complexities of SL-AD and SL-FD are smaller than AL-AD and AL-FD, both SL-AD and SL-FD require repeating the training phase multiple times in proportion to the number of subcarriers or antennas, which significantly increases the computation time. These results are consistent with the time overhead in Table \ref{tab: Table3}.

\section{CONCLUSION}\label{Conclusion}
This paper proposed a novel ML-based channel prediction approach for the wideband massive MIMO system that can be performed with small overhead for online re-training framework. To train the neural network with the small training time overhead, the concept of splitting and aggregating of training data is applied in the AL-AD and AL-FD. Especially, by defining the new type of channel form from the MIMO-OFDM system, i.e., a frequency domain channel, AL-FD can provide additional prediction performance gain compared to AL-AD.
The numerical results showed that AL-FD has superior channel prediction performance than AL-AD, SL-AD, SL-FD, SL-AD and SL-FD with FLIP and MAML, and JL, despite requiring only a small amount of time slots for training data collection, and we also analyzed various correlation properties to justify the numerical results.
We believe utilization of the idea of AL approach and the frequency domain channel can improve the performance of various wireless communication systems that rely on ML. For example, exploring the integration of ML techniques such as meta-learning into the AL approach offers a promising avenue, potentially combining strengths to enhance prediction accuracy in dynamic wireless environments.

\bibliographystyle{IEEEtran}
\bibliography{Reference}

\begin{thebibliography}{10}
\providecommand{\url}[1]{#1}
\csname url@samestyle\endcsname
\providecommand{\newblock}{\relax}
\providecommand{\bibinfo}[2]{#2}
\providecommand{\BIBentrySTDinterwordspacing}{\spaceskip=0pt\relax}
\providecommand{\BIBentryALTinterwordstretchfactor}{4}
\providecommand{\BIBentryALTinterwordspacing}{\spaceskip=\fontdimen2\font plus
\BIBentryALTinterwordstretchfactor\fontdimen3\font minus \fontdimen4\font\relax}
\providecommand{\BIBforeignlanguage}[2]{{%
\expandafter\ifx\csname l@#1\endcsname\relax
\typeout{** WARNING: IEEEtran.bst: No hyphenation pattern has been}%
\typeout{** loaded for the language `#1'. Using the pattern for}%
\typeout{** the default language instead.}%
\else
\language=\csname l@#1\endcsname
\fi
#2}}
\providecommand{\BIBdecl}{\relax}
\BIBdecl

\bibitem{Larsson2014}
E.~G. Larsson, O.~Edfors, F.~Tufvesson, and T.~L. Marzetta, ``Massive {MIMO} for next generation wireless systems,'' \emph{IEEE Commun. Mag.}, vol.~52, no.~2, pp. 186--195, Feb. 2014.

\bibitem{Marzetta2010}
T.~L. Marzetta, ``Noncooperative cellular wireless with unlimited numbers of base station antennas,'' \emph{IEEE Trans. Wireless Commun.}, vol.~9, no.~11, pp. 3590--3600, Nov. 2010.

\bibitem{Ramya2009}
T.~R. Ramya and S.~Bhashyam, ``Using delayed feedback for antenna selection in {MIMO} systems,'' \emph{IEEE Trans. Wireless Commun.}, vol.~8, no.~12, pp. 6059--6067, Dec. 2009.

\bibitem{Papa2017}
A.~K. Papazafeiropoulos, ``Impact of general channel aging conditions on the downlink performance of massive {MIMO},'' \emph{IEEE Trans. Veh. Technol.}, vol.~66, no.~2, pp. 1428--1442, Feb. 2017.

\bibitem{Truong2013}
K.~T. Truong and R.~W. Heath, ``Effects of channel aging in massive {MIMO} systems,'' \emph{J. Commun. Netw.}, vol.~15, no.~4, pp. 338--351, Aug. 2013.

\bibitem{Kong2015}
C.~Kong, C.~Zhong, A.~K. Papazafeiropoulos, M.~Matthaiou, and Z.~Zhang, ``Sum-rate and power scaling of massive {MIMO} systems with channel aging,'' \emph{IEEE Trans. Commun.}, vol.~63, no.~12, pp. 4879--4893, Dec. 2015.

\bibitem{Qin2022}
Z.~Qin, H.~Yin, Y.~Cao, W.~Li, and D.~Gesbert, ``A partial reciprocity-based channel prediction framework for {FDD} massive {MIMO} with high mobility,'' \emph{IEEE Trans. Wireless Commun.}, vol.~21, no.~11, pp. 9638--9652, Nov. 2022.

\bibitem{Turan2024}
\BIBentryALTinterwordspacing
N.~Turan, B.~Böck, K.~J. Chan, B.~Fesl, F.~Burmeister, M.~Joham, G.~Fettweis, and W.~Utschick, ``Wireless channel prediction via {Gaussian} mixture models,'' in \emph{arXiv:2402.08351}, Feb. 2024. [Online]. Available: \url{https://arxiv.org/abs/2402.08351}
\BIBentrySTDinterwordspacing

\bibitem{Wang2023}
L.~Wang, G.~Liu, J.~Xue, and K.-K. Wong, ``Channel prediction using ordinary differential equations for {MIMO} systems,'' \emph{IEEE Trans. Veh. Technol.}, vol.~72, no.~2, pp. 2111--2119, Feb. 2023.

\bibitem{MLP}
H.~Kim, S.~Kim, H.~Lee, C.~Jang, Y.~Choi, and J.~Choi, ``Massive {MIMO} channel prediction: Kalman filtering vs. machine learning,'' \emph{IEEE Trans. Commun.}, vol.~69, no.~1, pp. 518--528, Jan. 2021.

\bibitem{RNN1}
W.~Jiang and H.~D. Schotten, ``Neural network-based fading channel prediction: A comprehensive overview,'' \emph{IEEE Access}, vol.~7, pp. 118\,112--118\,124, Aug. 2019.

\bibitem{RNN2}
------, ``Recurrent neural networks with long short-term memory for fading channel prediction,'' in \emph{Proc. IEEE 91st Veh. Technol. Conf. (VTC2020-Spring)}, May 2020, pp. 1--5.

\bibitem{RNN3}
M.~K. Shehzad, L.~Rose, S.~Wesemann, and M.~Assaad, ``{ML}-based massive {MIMO} channel prediction: Does it work on real-world data?'' \emph{IEEE Wireless Commun. Lett.}, vol.~11, no.~4, pp. 811--815, Apr. 2022.

\bibitem{Shehzad2022}
M.~K. Shehzad, L.~Rose, M.~F. Azam, and M.~Assaad, ``Real-time massive {MIMO} channel prediction: A combination of deep learning and {NeuralProphet},'' in \emph{GLOBECOM 2022 - IEEE Global Commun. Conf.}, 2022, pp. 1423--1428.

\bibitem{Hao2022}
H.~Jiang, M.~Cui, D.~W.~K. Ng, and L.~Dai, ``Accurate channel prediction based on transformer: Making mobility negligible,'' \emph{IEEE J. Sel. Areas Commun.}, vol.~40, no.~9, pp. 2717--2732, Sep. 2022.

\bibitem{CNN2}
J.~Yuan, H.~Q. Ngo, and M.~Matthaiou, ``Machine learning-based channel prediction in massive {MIMO} with channel aging,'' \emph{IEEE Trans. Wireless Commun.}, vol.~19, no.~5, pp. 2960--2973, May 2020.

\bibitem{BNN}
Z.~Tao and S.~Wang, ``Improved downlink rates for {FDD} massive {MIMO} systems through {Bayesian} neural networks-based channel prediction,'' \emph{IEEE Trans. Wireless Commun.}, vol.~21, no.~3, pp. 2122--2134, Mar. 2022.

\bibitem{OFDM1}
C.~Wu, X.~Yi, Y.~Zhu, W.~Wang, L.~You, and X.~Gao, ``Channel prediction in high-mobility massive {MIMO}: From spatio-temporal autoregression to deep learning,'' \emph{IEEE J. Sel. Areas Commun.}, vol.~39, no.~7, pp. 1915--1930, Jul. 2021.

\bibitem{OFDM2}
G.~Liu, Z.~Hu, L.~Wang, J.~Xue, H.~Yin, and D.~Gesbert, ``Spatio-temporal neural network for channel prediction in massive {MIMO-OFDM} systems,'' \emph{IEEE Trans. Commun.}, vol.~70, no.~12, pp. 8003--8016, Dec. 2022.

\bibitem{OFDM3}
B.~Ko, H.~Kim, and J.~Choi, ``Machine learning-based channel prediction exploiting frequency correlation in massive {MIMO} wideband systems,'' in \emph{2021 Int. Conf. Inf. Commun. Technol. Converg. (ICTC)}, Oct. 2021, pp. 1069--1071.

\bibitem{Kim2023}
H.~Kim, J.~Choi, and D.~J. Love, ``Massive {MIMO} channel prediction via meta-learning and deep denoising: {Is} a small dataset enough?'' \emph{IEEE Trans. Wireless Commun.}, vol.~22, no.~12, pp. 9278--9290, Dec. 2023.

\bibitem{Training}
X.~Ying, ``An overview of overfitting and its solutions,'' \emph{J. Phys.: Conf. Ser.}, vol. 1168, no.~2, p. 022022, Feb. 2019.

\bibitem{DATAAUG1}
C.~Shorten and T.~M. Khoshgoftaar, ``A survey on image data augmentation for deep learning,'' \emph{J. Big Data}, vol.~6, no.~1, pp. 1--48, Jul. 2019.

\bibitem{DATAAUG2}
A.~Mumuni and F.~Mumuni, ``Data augmentation: {A} comprehensive survey of modern approaches,'' \emph{Array}, vol.~16, p. 100258, Dec. 2022.

\bibitem{DATAAUG3}
\BIBentryALTinterwordspacing
W.~Qingsong, S.~Liang, Y.~Fan, S.~Xiaomin, G.~Jingkun, W.~Xue, and X.~Huan, ``Time series data augmentation for deep learning: {A} survey,'' in \emph{arXiv:2002.12478}, Aug. 2021. [Online]. Available: \url{https://arxiv.org/abs/2002.12478}
\BIBentrySTDinterwordspacing

\bibitem{Timothy2022}
T.~Hospedales, A.~Antoniou, P.~Micaelli, and A.~Storkey, ``Meta-learning in neural networks: {A} survey,'' \emph{IEEE Trans. Pattern Anal. Mach. Intell.}, vol.~44, no.~9, pp. 5149--5169, Sep. 2022.

\bibitem{Offline1}
Y.~Zhang, Y.~Wu, A.~Liu, X.~Xia, T.~Pan, and X.~Liu, ``Deep learning-based channel prediction for {LEO} satellite massive {MIMO} communication system,'' \emph{IEEE Wireless Commun. Lett.}, vol.~10, no.~8, pp. 1835--1839, Aug. 2021.

\bibitem{Offline2}
Y.~Liao, Y.~Hua, and Y.~Cai, ``Deep learning based channel estimation algorithm for fast time-varying {MIMO-OFDM} systems,'' \emph{IEEE Commun. Lett.}, vol.~24, no.~3, pp. 572--576, Mar. 2020.

\bibitem{ADAM}
\BIBentryALTinterwordspacing
D.~P. Kingma and J.~Ba, ``Adam: {A} method for stochastic optimization,'' in \emph{arXiv:1412.6980}, Jan. 2017. [Online]. Available: \url{http://arxiv.org/abs/1412.6980}
\BIBentrySTDinterwordspacing

\bibitem{Diversity}
Z.~Gong, P.~Zhong, and W.~Hu, ``Diversity in machine learning,'' \emph{IEEE Access}, vol.~7, pp. 64\,323--64\,350, May 2019.

\bibitem{Redundancy}
\BIBentryALTinterwordspacing
V.~Birodkar, H.~Mobahi, and S.~Bengio, ``Semantic redundancies in image-classification datasets: {The} 10\% you don't need,'' Jan. 2019. [Online]. Available: \url{https://arxiv.org/abs/1901.11409}
\BIBentrySTDinterwordspacing

\bibitem{Finn2017}
C.~Finn, P.~Abbeel, and S.~Levine, ``Model-agnostic meta-learning for fast adaptation of deep networks,'' in \emph{Proc. 34th Int. Conf. Mach. Learn.}, 2017, pp. 1126--1135.

\bibitem{ConvLSTM}
X.~Shi, Z.~Chen, H.~Wang, D.-Y. Yeung, W.-K. Wong, and W.-C. Woo, ``Convolutional {LSTM} network: {A} machine learning approach for precipitation nowcasting,'' in \emph{Proc. 28th Int. Conf. Neural Inf. Process. Syst.}, 2015, pp. 802--810.

\bibitem{QUA}
\emph{Quasi Deterministic Radio Channel Generator User Manual and Documentation}, {Fraunhofer Heinrich Hertz Institute Wireless Communications and Networks}, Jul. 2021.

\bibitem{Tao2021}
Z.~Tao and S.~Wang, ``How often do we need to estimate wireless channels in massive {MIMO} with channel aging?'' in \emph{GLOBECOM 2021 - IEEE Global Commun. Conf.}, 2021, pp. 1--6.

\end{thebibliography}

\end{document}